\documentclass[%
 reprint,
 amsmath,amssymb,
 aps,
floatfix,
]{revtex4-2}

\usepackage{graphicx}
\usepackage{dcolumn}
\usepackage{bm}
\usepackage{chngcntr}
\usepackage{float}
\usepackage{hyperref}
\usepackage{ulem}

\usepackage{color}
\usepackage[cmyk]{xcolor}

\makeatletter
\newcommand*{\rom}[1]{\expandafter\@slowromancap\romannumeral #1@}
\makeatother

\begin{document}

\preprint{APS/123-QED}

\title{Identifying the impact of local connectivity patterns on dynamics in excitatory-inhibitory networks}%
\author{Yuxiu Shao}
\email{ivyerosion@gmail.com}
\affiliation{School of Systems Science, Beijing Normal University, Beijing, China\\
Laboratoire de Neurosciences Cognitives et Computationnelles, INSERM U960, Ecole Normale Superieure - PSL Research University, Paris, France
}
\author{David Dahmen}%
\affiliation{%
 Institute for Advanced Simulation (IAS-6) Computational and Systems Neuroscience, Jülich Research Center, Jülich, Germany
}%
\author{Stefano Recanatesi}%
\affiliation{%
 Technion, Israel Institute of Technology, Haifa, Israel
}%
\author{Eric Shea-Brown}
 \affiliation{Department of Applied Mathematics and
Computational Neuroscience Center, University of Washington, Seattle, WA, USA\\
Allen Institute for Brain Science, Seattle, WA, USA
}
\author{Srdjan Ostojic}
\email{srdjan.ostojic@ens.fr}
\affiliation{Laboratoire de Neurosciences Cognitives et Computationnelles, INSERM U960, Ecole Normale Superieure - PSL Research University, Paris, France
}

\date{\today}

\begin{abstract}
Networks of excitatory and inhibitory (EI) neurons  form a canonical circuit in the brain. Seminal theoretical results on dynamics of such networks are based on the assumption that synaptic strengths depend on the type of neurons they connect, but are otherwise statistically independent. 
Recent synaptic physiology datasets however highlight the prominence of specific connectivity patterns that go well beyond what is expected from independent connections.  
While decades of influential research have demonstrated the strong role of the basic EI cell type structure, to which extent additional connectivity features influence dynamics remains to be fully determined. 
Here we examine the effects of pair-wise connectivity motifs on the linear dynamics in excitatory-inhibitory networks using an analytical framework that approximates the connectivity in terms of low-rank structures. This low-rank approximation is based on a mathematical derivation of the dominant eigenvalues of the connectivity matrix, and predicts the impact on responses to external inputs of connectivity motifs and their interactions with cell-type structure.
Our results reveal that a particular pattern of connectivity, chain motifs, have a much stronger impact on dominant eigenmodes than other pair-wise motifs. In particular,
an over-representation of chain motifs induces a strong positive eigenvalue in inhibition-dominated networks and generates a potential instability that requires revisiting the classical excitation-inhibition balance criteria. Examining effects of external inputs, we show that chain motifs can on their own induce paradoxical responses, where an increased input to inhibitory neurons leads to a decrease in their activity due to the recurrent feedback. These findings have direct implications for the interpretation of experiments in which responses to optogenetic perturbations are measured and used to infer the dynamical regime of cortical circuits.  
\end{abstract}

\maketitle
\section*{Introduction}

Circuits of excitatory and inhibitory (EI) neurons are believed to form the fundamental components of information-processing  in the brain \cite{douglas2004neuronal,harris2013cortical,harris2015neocortical,luo2021architectures}. Network models of recurrently-connected excitatory and inhibitory units have therefore become an essential tool for understanding  neural dynamics and computation. Such models have helped uncover fundamental principles such as the role of excitation-inhibition balance for sustaining irregular activity \cite{shadlen1994noise,shadlen1998variable,van1996chaos, amit1997model,hennequin2017inhibitory}, and the importance of inhibition for stabilizing neural activity \cite{amit1997model, tsodyks1997paradoxical, ozeki2009inhibitory, litwin2016inhibitory} and normalizing responses \cite{rubin2015stabilized}. A phenomenon that has attracted particular attention are {\it paradoxical responses}, which refer to situations where an increase in the external input to the inhibitory neurons results in a decrease of their activity because of recurrent interactions \cite{tsodyks1997paradoxical, ozeki2009inhibitory, ahmadian2013analysis,rubin2015stabilized,miller2020generalized,sadeh2021inhibitory, rubin2015stabilized,litwin2016inhibitory}.
Recent theoretical analyses have argued that such paradoxical responses to external inputs can reveal the dynamical regime of the underlying excitatory-inhibitory network \cite{ahmadian2013analysis,hennequin2018dynamical,ahmadian2021dynamical}, and these insights have been used to interpret experimental measurements of responses to optogenetic perturbations \cite{sanzeni2020inhibition, palmigiano2020common, sanzeni2023mechanisms}.

\begin{figure}[ht]
\includegraphics[width=0.5\textwidth]{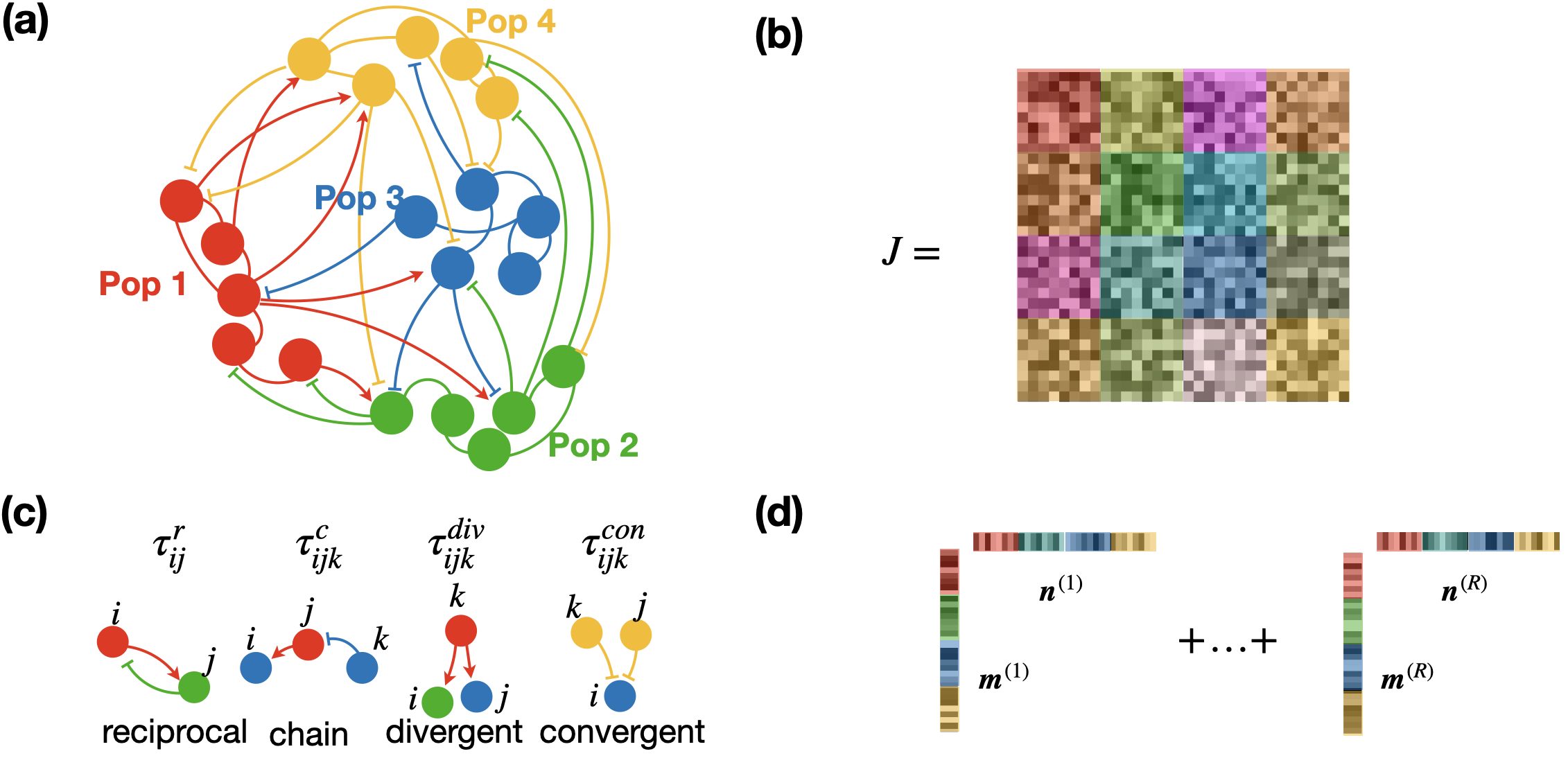}
\caption{\label{fig:Schematic} 
Schematic of the multi-population network model. 
(a) The network consists of $P$ populations, each represented by a different color. This population structure defines the statistics of synaptic connectivity. 
(b) The corresponding connectivity matrix consists of $P^2$ blocks. Synaptic weights within each block share identical statistical properties. 
(c) Pairs of synapses that share a neuron can form four different types of second-order motifs. The prevalence of each motif with respect to chance is quantified by a corresponding  pair-wise correlation coefficient.
(d) A low-rank approximation of the connectivity matrix can integrate both the population structure and the pair-wise motif statistics. }
\end{figure}

These seminal theoretical results on excitatory-inhibitory networks however are derived via a key simplifying assumption. In standard models, the strength of the synaptic coupling between any two neurons depends on their types, but is otherwise assumed to be an independent random variable uncorrelated across synapses. This assumption is typically used to reduce a full network to a simpler circuit model that describes how the mean activities of different populations interact through averaged synaptic weights  (Fig~\ref{fig:Schematic}(a,b).
Recent synaptic-resolution experimental datasets from various species and brain areas have however revealed the prevalence of non-trivial connectivity patterns \cite{winding2023connectome,campagnola2022local,pfeffer2013inhibition,lefort2009excitatory,hofer2011differential,peng2019high,rupprecht2018precise, song2005highly,dahmen2020strong}.
In particular, a recently released dataset from mice and humans \cite{campagnola2022local,dahmen2020strong} reported the prominence of second-order motifs - specific correlations between pairs of synapses, such as reciprocal, chain, convergent, and divergent motifs (Fig~\ref{fig:Schematic}(c))  - that go well beyond what is expected from independent connections, and highlights the need for a theoretical understanding of the effects of such patterns.
The presence of reciprocal connectivity motifs has been long recognized and examined within network models \cite{song2005highly,marti2018correlations,dahmen2020strong, hu2022spectrum, shao2023relating}. Other studies have argued for the importance of several types of synaptic motifs working together to determine statistical properties of network activity, such as average synchrony or correlation among neurons and network-wide dimensionality \cite{zhao2011synchronization, hu2018feedback,hu2022spectrum,dahmen2020strong,trousdale2012impact,hu2013motif,hu2014local,ocker2017linking,recanatesi19dimension}.  In a general theoretical analysis \cite{hu2018feedback}, chain motifs
-- corresponding to a pattern where neurons with stronger inputs also have stronger outputs --
were found to have a dominant role in determining population-averaged responses of networks to their inputs.  Nevertheless, the impact of chain motifs on excitation-inhibition balance -- including central issues of stability and paradoxical responses -- as well as the general interplay among these motifs and structures established by cell-type specific connectivity, remain open and intriguing questions.

In this study, we examine how the interaction between population structure and synaptic motifs influences recurrent dynamics and the presence of paradoxical responses.
To this end, we expand a previously introduced theoretical framework that allows us to reduce large networks of multiple populations to a low-dimensional description that incorporates connectivity statistics beyond the mean via a low-rank approximation \cite{shao2023relating}. Applying this theory to excitatory-inhibitory networks with chain motifs, we demonstrate that these connectivity patterns significantly impact the eigen-spectrum of the connectivity matrix, and thereby the overall recurrent dynamics. Specifically, an over-representation of chain motifs creates strong positive feedback even in inhibition-dominated networks, and therefore leads to a potential instability that requires revisiting the classical conditions of excitation-inhibition balance. Moreover, we found that chain motifs strongly influence the responses of different populations to external inputs, and can control whether the responses are paradoxical or not. 
We show that the paradoxical responses can be equivalently predicted from two approaches: (i) a low-rank approximation based on the outliers in the eigenspectrum;  (ii) an effective connectivity matrix which combines average synaptic weights and the strength of chain motifs. Altogether, our findings highlight the intricacy of the relationship between the responses to inputs and the underlying connectivity, and in particular sound a note of caution for interpreting results of experiments in terms of only average connectivity strengths among E and I cells.

This manuscript is organized as follows. {\bf Sec.~\ref{secs:Networkmodel}} defines the connectivity and the network model. {\bf Sec.~\ref{secs:LowrankApproach}} introduces the general theoretical framework based on a low-rank approximation of the connectivity, which allows us to analytically investigate the effects of local motifs, in particular chain and reciprocal motifs, on recurrent dynamics and responses to external inputs.
{\bf Secs.~\ref{secs:EigenvaluesandBalancedregime} and~\ref{secs:EigenvectorDynamics}} apply this theory to fully-connected and sparse excitatory-inhibitory networks with chain motifs. In {\bf Sec.~\ref{secs:EigenvaluesandBalancedregime}} we analyze the influence of chain motifs on the eigenspectra of these models. In {\bf Sec.~\ref{secs:EigenvectorDynamics}}, we use these results to study the responses to external inputs and under what condition they are paradoxical. 

\begin{table*}
    \caption{\label{tab:notation}List of notations.}
    \begin{ruledtabular}
    \begin{tabular}{lllll}
    \textrm{Notation}&
    \multicolumn{1}{l}{\textrm{Description}}\\
    \colrule
    $i,j$ & Single neuron indices\\
    $p,q$ & Population indices\\ 
    $N_p$ & Number of neurons in population $p$\\
    $\alpha_p$ & Fraction of neurons in population $p$\\
    $J^0_{pq}$ & Mean synaptic weight from population $p$ to population $q$\\
    $\sigma_{pq}$ & standard deviation of the synaptic weights from population $p$ to population $q$\\
    $\sigma$ & Re-scaled homogeneous standard deviation of the synaptic weights\\ 
    $\tau^{c/r}_{pq}$ & Correlation coefficient of the chain/reciprocal  connectivity motifs\\
    $c$ & Connection probability in sparse networks\\
    $J$ & Excitatory synaptic weights in the sparse E-I network\\
    $g$ & Relative (mean) strength of inhibitory to excitatory synapses in (Gaussian) sparse E-I network  \\
    $\gamma$ & Ratio of inhibitory to excitatory population
    size\\
\end{tabular}
\end{ruledtabular}
\end{table*}

\section{\label{secs:Networkmodel}Network model}
\subsection{Network connectivity}
We consider networks of $N$ recurrently connected neurons. The connectivity is represented by a  matrix $\mathbf{J}$, where the entry $J_{ij}$ corresponds to the synaptic weight from neuron $j$ to neuron $i$. The statistics of network connectivity are then fully described by the joint distribution $\mathcal{P}(\{J_{ij}\})$ of the $N^2$ synaptic weights. 

We assume that the network consists of $P$ non-overlapping populations that determine the connectivity statistics as follows. 
The connectivity matrix $\mathbf{J}$  has a $P \times P$ block  structure defined by the populations, where synaptic weights within a specific block share identical statistical properties (Fig~\ref{fig:Schematic}(b)). We denote as $N_p$ the number of neurons in population $p$ and $\alpha_p=\frac{N_p}{N}$ the corresponding fraction. The main notations are summarized in Table \ref{tab:notation}.

Following a common approach \cite{song2005highly,nykamp2005revealing,zhao2011synchronization,trousdale2012impact,hu2013motif,hu2014local,ocker2017linking,hu2022spectrum}, we
begin by characterizing connectivity through the marginal distributions $\mathcal{P}(J_{ij}=J)$ of individual synaptic weights, and progressively incorporate higher-order correlations. 
In this study, we consider the first and second orders, that is, the marginal distribution of individual synaptic weights and the pair-wise correlations. We denote as {\it motifs}  correlations between weights of pairs of synapses that share a neuron (Fig~\ref{fig:Schematic}(c)).
We specifically focus on {\it chain motifs} corresponding to correlations
 between synapses $J_{ij}$  and $J_{jk}$ which share the intermediate neuron $j$, but  $i\neq k$. The strength of chain motifs can be quantified by the correlation coefficient 
\begin{equation}\label{eq:definition_chain_main}
\tau^c_{ijk}=\frac{\left[ J_{ij} J_{jk} \right] - \left[ J_{ij}\right] \left[J_{jk} \right]}{\sqrt{\left[ (J_{ij}-[J_{ij}])^2\right]\left[(J_{jk}-[J_{jk}])^2\right]}}.
\end{equation}

This measure of motif strength takes the form of a Pearson correlation coefficient, representing the prevalence of correlations between a pair of synaptic weights compared to the  independent case.
Here, and throughout this study, square brackets denote the average over the full connectivity distribution.

We contrast the effects of chain motifs with reciprocal motifs corresponding to correlations $\tau^r_{ij}$ between connections $J_{ij}$ and $J_{ji}$ \cite{shao2023relating}:
\begin{equation}\label{eq:definition_reciprocal_main}
\tau^r_{ij}=\frac{\left[ J_{ij} J_{ji} \right] - \left[ J_{ij}\right] \left[J_{ji} \right]}{\sqrt{\left[ (J_{ij}-[J_{ij}])^2\right]\left[(J_{ji}-[J_{ji}])^2\right]}}.
\end{equation}
Pairs of synapses sharing a neuron can form two additional types of correlations corresponding to convergent and divergent motifs. We show in Appendix.~\ref{ap:GeneralMotifs} that, in the limit of large networks, they do not contribute to the dynamics we study here.

Because of the assumed population structure, the marginal distributions $\mathcal{P}(J_{ij})$, and the  correlation coefficients  $\tau^c_{ijk}$ and $\tau^r_{ij}$ depend only on the populations $p,q$ and $s$ to which the neurons $i$, $j$ and $k$ belong, so that
\begin{equation}\label{eq:ProbDefinition}
\begin{aligned}
\mathcal{P}(J_{ij}=J) &= f^{pq}(J),\\
\tau^c_{ijk} &=\tau^c_{pqs},\\
\tau^r_{ij} &= \tau^r_{pq}.
\end{aligned}
\end{equation}
Here $p,q,s\in \{1\dots P\}$ (Fig~\ref{fig:Schematic}(a,~b)), and $f^{pq}(J)$ is the marginal distribution of synaptic weights from population $p$ to population $q$. 

The connectivity matrix $\mathbf{J}$ can in general be split into a sum of its mean and random parts
\begin{equation}\label{eq:LocallyDefined}
\mathbf{J} = \mathbf{J^0}+\mathbf{Z}
\end{equation}
where $J^0_{ij}=\left[ J_{ij}\right]$ are mean connectivity weights and $z_{ij}=J_{ij}-\left[ J_{ij}\right]$  the zero-mean random part \cite{shao2023relating}. 

These two parts are  characterized by statistics
\begin{equation}\label{eq:LocallyDefinedStats}
\begin{aligned}
J^0_{ij}&=J^0_{pq}\\
[z_{ij}^2]&=\sigma_{{pq}}^2\\
\frac{\left[z_{ij}z_{jk}\right]}{\sqrt{[z_{ij}^2][z_{jk}^2]}}&=\tau^c_{pqs}\\
\frac{\left[z_{ij}z_{ji}\right]}{\sqrt{[z_{ij}^2][z_{ji}^2]}}&=\tau^r_{pq}
\end{aligned}
\end{equation}
where neurons with indexes $i,~j,~k$ belong to populations $p,~q$ and $s$ respectively 
and $i\neq k$. The mean matrix $\mathbf{J^0}$ therefore has a $P \times P$ block structure, where entries within each block are identical, so that its rank is $R_0\leq P$. In contrast, the random component $\mathbf{Z}$ is in general of rank $N$. The statistics of its entries however exhibit a block-like structure, where variances  $\sigma_{pq}^2$ and reciprocal correlation coefficients $\tau^r_{pq}$ define $P \times P$ matrices, while the chain correlation coefficients $\tau_{c,pqs}$ define a $P\times P\times P$ tensor. Both $\tau^c_{pqs}$ and $\tau^r_{pq}$ range from $-1$ to $1$. We further simplify the parameters by assuming that the chain correlation coefficient depends only on the populations of the pre-synaptic neurons of the two synapses, consequently, $\tau^c_{pqs} = \tau^c_{qs}$. 

\subsection{Network dynamics}
Our goal is to examine the effects of local connectivity statistics on the steady-state response to external inputs. We therefore focus on a simple linear network of rate units, where the firing rate $r_i$ of unit $i$ obeys:
\begin{equation}\label{eq:NetworkDynamics}
\tau\frac{d}{dt}r_i(t) = -r_i(t) + \sum_{j=1}^N J_{ij}r_j(t) + I_i^{ext}(t).
\end{equation}
Here, $-r_i(t)$ is a standard leak term,  the second term on the r.h.s. is the recurrent input from other units in the network, and the third term is an external input.

The firing rate dynamics of neurons in vector format can be written as
\begin{equation}\label{eq:ExternalDynamics}
\begin{aligned}
    \tau\frac{d}{dt} \boldsymbol{r}(t) = -\boldsymbol{r}(t)+\mathbf{J}\boldsymbol{r}(t)+\boldsymbol{I}^{ext}(t),
\end{aligned}
\end{equation}
so that the steady-state activity is given by
\begin{equation}\label{eq:EquilibriumDynsInverse}
    \boldsymbol{r}^* = \big(\mathbf{1}-\mathbf{J}\big)^{-1}\boldsymbol{I}^{ext}.
\end{equation}
Here $\mathbf{1}$ denotes the $N\times N$ identity matrix.

Our aim is to determine how external inputs to different populations affect the steady-state activity. To this end, we will examine the response matrix $\boldsymbol{\chi}$, where the entry $\chi_{ij}$  represents the change of the steady-state activity of neuron $i$ resulting from a change in the input current to neuron $j$:
\begin{equation}
\chi_{ij} = \frac{d r_i}{dI^{ext}_j}.
\end{equation}
From Eq.~\eqref{eq:EquilibriumDynsInverse} we have 
\begin{equation}
\boldsymbol{\chi} = (\mathbf{1}-\mathbf{J})^{-1}. \label{eq:response_full_rank}
\end{equation}

The diagonal entries of the response matrix, denoted as $\chi_{ii}$, represent the activity change of neuron  $i$ in response to the direct external input they receive. When this term is negative, this indicates that the neuron increases (decreases) its firing rate with a decrease (increase) in the external excitatory input. Generalizing the notion of paradoxical responses of inhibitory neurons \cite{tsodyks1997paradoxical}, 
we denote any negative diagonal element of the response function as a {\it paradoxical} response.

To examine the relationship between local connectivity statistics and paradoxical responses to external inputs, we compare two approaches: (i) responses obtained from a low-rank approximation based on the dominant eigenmodes of the random connectivity matrix $\mathbf{J}$ \citep{shao2023relating}; (ii) a direct approximation of the average response function in which the random connectivity matrix $\mathbf{J}$ is replaced by a deterministic matrix  $\mathbf{J}^{eff}$ that combines the first and second order statistics \citep{hu2018feedback}.

\section{\label{secs:LowrankApproach}Low-rank approximation}
To determine how  connectivity 
shapes the response function of the network,
we perform a low-rank approximation of the  connectivity matrix $\mathbf{J}$ (Fig~\ref{fig:Schematic}(d)):

\begin{eqnarray}\label{eq:low-rank-approx}
\mathbf{J}&\approx&\frac{1}{N}\sum_{r=1}^{R}\boldsymbol{m}^{(r)}\boldsymbol{n}^{(r)\intercal}\\
&=&\frac{1}{N}\mathbf{M}\mathbf{N}^{\intercal}
\end{eqnarray}

Here $\boldsymbol{m}^{(r)}$ and $\boldsymbol{n}^{(r)}$ for $r=1
\ldots R$ are $N$-dimensional vectors, and $\mathbf{M}$ and $\mathbf{N}$ are $N\times R$ matrices obtained by concatenating these vectors.

While there exists a variety of methods for forming a low-rank approximation, here we use a simple truncated eigen-decomposition, 
which assumes that the connectivity matrix 
$\mathbf{J}$ is diagonalizable.
Specficially, we take the vectors $\boldsymbol{n}^{(r)},~\boldsymbol{m}^{(r)}\in \mathbb{R}^{N}$ to be respectively the  $r-$th left and right eigenvectors $\boldsymbol{L}_r$ and $\boldsymbol{R}_r$ re-scaled by $\sqrt{N}$: 
\begin{equation}\label{eq:normalization0}
\boldsymbol{m}^{(r)}=\sqrt{N}\boldsymbol{R}_r,\,\,\,\boldsymbol{n}^{(r)}=\lambda_r\sqrt{N}\boldsymbol{L}_r.
\end{equation}
Here, the eigenvectors are ordered by the decreasing  absolute value of their eigenvalue $\lambda_r$, and obey

\begin{equation}\label{eq:normalization1}
\begin{aligned}
\mathbf{J}\boldsymbol{R}_r=\boldsymbol{R}_r\lambda_r,\,\,\,
\boldsymbol{L}_r^{\intercal}\mathbf{J}=\lambda_r\boldsymbol{L}_r^{\intercal},\,\,\,
\boldsymbol{L}_r^{\intercal}\boldsymbol{R}_{r'}=\delta_{rr'}.
\end{aligned}
\end{equation}
Additionally, we impose the constraint that $\boldsymbol{R}_r$ has unit norm.
By retaining only the first $R$ eigenmodes  we obtain a rank-$R$ approximation that preserves the top $R$ eigenvalues of $\mathbf{J}$.

From Eq.~\eqref{eq:EquilibriumDynsInverse}, this leads to a low-rank approximation of the steady state activity
\begin{equation}\label{eq:steady_state_low_rank}
    \begin{aligned}
        \boldsymbol{r}^* = \big(\mathbf{1}-\mathbf{M}\mathbf{N}^{\intercal}/N\big)^{-1}\boldsymbol{I}^{ext}.
    \end{aligned}
\end{equation}
Using the Woodbury matrix identity, the steady-state activity can be expressed as:
\begin{equation}
    \begin{aligned}
        \boldsymbol{r}^* = \big(\mathbf{1}+\mathbf{M}(\mathbf{1}-\mathbf{\Lambda})^{-1}\mathbf{N}^{\intercal}/N\big)\boldsymbol{I}^{ext},
    \end{aligned}
\end{equation}
where $\mathbf{\Lambda}$ is the $R\times R$ diagonal matrix containing the top $R$ eigenvalues of $\mathbf{J}$.

In the low-rank approximation, the entries of the linear response matrix are therefore given by
\begin{equation}\label{eq:responseLowrank}
\chi_{ij} = \delta_{ij} + \frac{1}{N}\sum_{r=1}^R\frac{m_i^{(r)}n_j^{(r)}}{1 - \lambda_r},
\end{equation}
where $\delta_{ij}$ is the Kronecker delta. Note that in \cite{hu2018feedback}, the response function in networks with chain motifs was computed using a different method, based on a power series expansion and resumming of the matrix inverse in Eq.~\eqref{eq:response_full_rank}.

To identify the $R$ dominant eigenmodes of $\mathbf{J}$, we leverage prior results on the eigenspectra of matrices with a low-rank plus random structure similar to  Eq.\eqref{eq:LocallyDefined}. For such matrices, the eigenspectra typically consist of two components in the complex plane. One component is a continuously-distributed bulk determined by $\mathbf{Z}$, and the other is a set of discrete outliers controlled by the low-rank structure \cite{mastrogiuseppe2018linking,schuessler2020dynamics,logiaco2021thalamic,rajan2006eigenvalue,tao2013outliers,dahmen2019second}.
Previous works have examined the influence of local connectivity motifs on the eigenvalue bulk \cite{aljadeff2015transition,kuczala2016eigenvalue,dahmen2020strong,hu2022spectrum}. Here we instead focus on  eigenvalue outliers and their associated eigenvectors. Specifically, we take the rank $R$ in our low-rank approximation to be equal to  the number of  eigenvalue outliers.

In the subsequent section, we identify the impact of chain and reciprocal motifs on the outlying eigenvalues and corresponding eigenvectors, and then use the low-rank approximation to determine their influence on the steady state response of the network.

\section{\label{secs:EigenvaluesandBalancedregime}Eigenvalue outliers}
In this section, we study the impact of chain motifs on outlying eigenvalues of the connectivity matrix. We first outline the general theory for networks consisting of $P$ populations. We then apply it to excitatory-inhibitory networks consisting of two populations, and contrast fully-connected and sparse networks.

\subsection{General approach\label{sec:EigenvaluesMethod}}
 To investigate the impact of local motif statistics on eigenvalues, we expand on earlier work on random matrix theory \cite{tao2013outliers,schuessler2020dynamics,greenbaum2020first,shao2023relating,zhao2011synchronization,nykamp2005revealing}. The main steps of the mathematical deviations are outlined below, details of the derivation are provided in Appendix \ref{ap:EigenvalueCalculations}.

Our starting point is the fact that the mean connectivity matrix $\mathbf{J^0}$ of a network with $P$ populations consists of $P$ blocks and is therefore of rank $R_0\leq P$. Therefore, it can be exactly expressed as
\begin{equation}
\mathbf{J^0}=\frac{1}{N}\sum_{r=1}^{R_0}\boldsymbol{m}_0^{(r)}\boldsymbol{n}_0^{(r)\intercal} = \frac{1}{N}\mathbf{M}_0\mathbf{N}_0^{\intercal},
\end{equation}
where $\mathbf{M}_0$ and $\mathbf{N}_0$ are two $\mathbb{R}^{N\times R_0}$ matrices, and their $r-$th columns correspond to the right and left eigenvectors $\boldsymbol{m}_0^{(r)}$ and $\boldsymbol{n}_0^{(r)}$ of $\mathbf{J^0}$ associated with the non-zero eigenvalue $\lambda_r^0$. 
The full connectivity matrix $\mathbf{J}$ (Eq.\eqref{eq:LocallyDefined}) can therefore be written as
\begin{equation}\label{eq:MeanPlusRandom}
\mathbf{J} = \frac{1}{N}\mathbf{M}_0\mathbf{N}_0^{\intercal}+\mathbf{Z}.
\end{equation}
Any eigenvalue $\lambda$ of  $\mathbf{J}$ satisfies
\begin{equation} \label{eq:characteristic_polynomialJ}
    \det(\mathbf{J}-\mathbf{1}\lambda) = 0.
\end{equation}
After substituting Eq.~\eqref{eq:MeanPlusRandom} and applying the matrix determinant lemma, the determinant in Eq.~\ref{eq:characteristic_polynomialJ} can be expressed as

\begin{equation}\label{eq:determinantLemma}
\begin{aligned}
    \det(\mathbf{M}_0&\mathbf{N}_0^{\intercal}/N+\mathbf{Z}-\mathbf{1}\lambda)\\
    &= \det(\mathbf{Z}-\mathbf{1}\lambda)\det(\mathbf{1}+\mathbf{N}_0^{\intercal}(\mathbf{Z}-\mathbf{1}\lambda)^{-1}\mathbf{M}_0/N)\\
    &= \det(\mathbf{Z}-\mathbf{1}\lambda)\frac{1}{\lambda^N}\det\left(\lambda\mathbf{1}-\frac{1}{N}\mathbf{N}_0^{\intercal}(\mathbf{1}-\mathbf{Z}/\lambda)^{-1}\mathbf{M}_0\right).
\end{aligned}
\end{equation}

To apply the matrix determinant lemma, we assumed that $\mathbf{Z}-\mathbf{1}\lambda$ is invertible, so that the first term on the r.h.s of Eq.~\eqref{eq:determinantLemma} is nonzero. This is guaranteed if $\lambda$ is larger than the radius of the spectrum of $\mathbf{Z}$, which we refer to  as the {\it bulk}. 
The zeros of the second term on the r.h.s corresponds to potential eigenvalues generated by the interaction between $\mathbf{Z}$ and the mean connectivity $\mathbf{J^0}$.   We refer to these eigenvalues as {\it outliers} when they lie outside of the bulk spectrum. Note that, if the random component $\mathbf{Z}=\mathbf{0}$, the second term yields  the non-zero eigenvalues of the pure low-rank mean connectivity $\mathbf{J^0}$.

Expanding the matrix inverse $(\mathbf{1}-\mathbf{Z}/\lambda)^{-1}$ in the second term of Eq.~\eqref{eq:determinantLemma}, we introduce the matrix
\begin{equation}\label{eq:seriesRepresentation}
    \mathbf{Q} = \sum_{k=0}^{\infty}\mathbf{N}_0^{\intercal}\mathbf{Z}^k\mathbf{M}_0/(N\lambda^k).
\end{equation}
This expansion converges for eigenvalues $\lambda$ that are outside of the bulk  because the norm of $\mathbf{Z}/\lambda$ is smaller than unity in this case.
The outliers  are then given by the polynomial equation 
\begin{equation}\label{eq:OutlierInfinite}
    \det(\lambda\mathbf{1} -\mathbf{Q})=0.
\end{equation}

We focus on the realization-averaged $\lambda$,  involving the averaging of Eq.\eqref{eq:OutlierInfinite} and, consequently, the averaging of Eq.\eqref{eq:seriesRepresentation} over $\mathbf{Z}$, which we denote with square brackets.
As we consider networks with only second-order correlations between synaptic weights, $[\mathbf{Z}^k]$ for odd $k$ are zero, and only even terms remain non-zero. For even $k=2l$, we show that in the limit of large networks, $[\mathbf{Z}^{2l}]\to [\mathbf{Z}^{2}]^l$ , with higher powers functioning in a sub-dominant way and thus being negligible in the following calculations (see Appendix \ref{ap:GeneralMotifs}).  

Applying geometric sequence summation in Eq.~\eqref{eq:seriesRepresentation}, we get  
\begin{equation}\label{eq:OutlierSecondorder}
    \begin{aligned}
    \left[\mathbf{Q}\right] &=\frac{1}{N}\mathbf{N}_0^{\intercal}\left(\mathbf{1}-\frac{\left[\mathbf{Z}^2\right]}{\lambda^2}\right)^{-1}\mathbf{M}_0.
    \end{aligned}
\end{equation}

The element $i,j$ of $\left[\mathbf{Z}^2\right]$ is given by the pair-wise correlations between synapses, and  can be expressed as
 
\begin{equation}\label{eq:motifQuadraticZ}
\left[\sum_{k=1}^Nz_{ik}z_{kj}\right] =\left \{
\begin{matrix}
N\sum_{q=1}^P \alpha_q\sigma_{{pq}}\sigma_{{qs}}\tau^c_{qs},&i\neq j,\\
N\sum_{q=1}^P \alpha_q\sigma_{{pq}}\sigma_{{qp}}\tau^r_{qp},&i=j,\
\end{matrix}
\right.
\end{equation}
where neurons $i$ and $j$ belong to populations $p$ and $s$, and $\alpha_q$ represents the fraction of neurons in population $q$.

Eq.~\eqref{eq:motifQuadraticZ} shows that diagonal entries of $[\mathbf{Z}^2]$ are determined by the strength of reciprocal motifs, while the non-zero off-diagonal entries are determined by the strength of chain motifs. Thus, the matrix $[\mathbf{Z}^2]$ can be decomposed into a diagonal matrix $\mathbf{D}\in \mathbb{R}^{N\times N}$ and a matrix $\mathbf{O}\in\mathbb{R}^{N\times N}$  consisting of $P^2$ blocks with constant entries within blocks that are  determined by the strength of chain-motifs across the different populations: 
\begin{equation}\label{eq:ZsquareDecompose_Maintext}
\left[\mathbf{Z}^2\right]= \mathbf{D}+\mathbf{O}.
\end{equation}

We further  express this block matrix as $\mathbf{O}=\mathbf{U}_o\mathbf{V}_o^{\intercal}$, where $\mathbf{U}_o,\mathbf{V}_o\in \mathbb{R}^{N\times R_o}$, and $R_o$ represents the rank of $\mathbf{O}$ which is maximally the number of distinct neuron populations $P$. 
To compute the matrix inverse $(\mathbf{1}-[\mathbf{Z}^2]/\lambda^2)^{-1}$, we then apply the Woodbury matrix identity, resulting in:
\begin{equation}\label{eq:matrixInverse_main}
\begin{aligned}
\left(\mathbf{1}-\frac{\left[\mathbf{Z}^2\right]}{\lambda^2}\right)^{-1} =& \frac{1}{\lambda^2}\mathbf{A}^{-1}\mathbf{U}_o\left(\mathbf{1}_{R_o}-\frac{1}{\lambda^2}\mathbf{V}_o^{\intercal}\mathbf{A}^{-1}\mathbf{U}_o\right)^{-1}\\
&\cdot\mathbf{V}_o^{\intercal}\mathbf{A}^{-1}+\mathbf{A}^{-1}\\
\mathbf{A}^{-1} &= {\rm diag}\left(\left\{\frac{\lambda^2}{\lambda^2-D_{ii}}\right\}\right)\\
\end{aligned}
\end{equation}
where $\mathbf{1}_{R_o}$ is an $\mathbb{R}^{R_o\times R_o}$ identity matrix.
Substituting the matrix inverse Eq.~\eqref{eq:matrixInverse_main} into Eqs.~\eqref{eq:OutlierSecondorder} and\eqref{eq:OutlierInfinite} yields a polynomial equation for the eigenvalue outliers of the connectivity matrix $\mathbf{J}$ as a function of the first and second-order statistics of synaptic strengths (see Appendix \ref{ap:EigenvalueCalculations} for details).

\subsection{\label{subsecs:EigenvalueGauss}Fully connected excitatory-inhibitory networks}

\subsubsection{Definition}
We first apply our approach to fully connected networks with Gaussian-distributed synaptic strengths. We consider networks  consisting of two populations, an excitatory and an inhibitory one, which we denote with indices $p,q \in\{E,I\}$. Their respective sizes are $N_E=\alpha_E N$ and $N_I=\alpha_I N$. 

The marginal distribution $f^{pq}$ of synaptic strengths from population $q$ to population $p$ is given by:
\begin{equation}\label{eq:Gaussianmarginal}
f^{pq}=\mathcal{N}(J^0_{pq},\sigma_{{pq}}^2)
\end{equation}
where $p,q\in \{E,I\}$.
For simplicity, we set $J^0_{EE}=J^0_{IE}=J^0$ and $J^0_{EI}=J^0_{II}=-g J^0$, where $g$ is the relative strength of inhibitory synapses with respect to excitatory ones.

The mean synaptic connectivity $\mathbf{J^0}$ is then of  rank $R_0=1$, and can be expressed as $\mathbf{J^0}=\boldsymbol{m}_0\boldsymbol{n}_0^{\intercal}/N$, where
\begin{equation}\label{eq:meanstructures_Gaussian}
\begin{aligned}
\boldsymbol{m}_0 &= \left[1\dots\right]^{\intercal}\\
\boldsymbol{n}_0 &= \left[NJ^0\dots, -NgJ^0\dots\right]^{\intercal},
\end{aligned}
\end{equation}
$\mathbf{J^0}$ has a unique non-zero eigenvalue
\begin{equation} \label{eq:gauss_lambda0}
\begin{aligned}
\lambda_0&=N_EJ^0-N_IgJ^0\\
&=(\alpha_E-g\alpha_I)J^0N.
\end{aligned}
\end{equation}
In the following, we will assume that the network is {\it inhibition-dominated} \cite{hennequin2018dynamical,ahmadian2021dynamical,sanzeni2020inhibition,vogels2011inhibitory}, i.e. that $\alpha_E-g\alpha_I\leq 0$, so that $\lambda_0\leq 0$. We refer to $\lambda_0$ as the {\it unperturbed eigenvalue}.

We moreover consider a situation where the variance $\sigma_{{pq}}^2$ is identical across populations and strong, i.e. proportional to $1/N$, so that the variance of the total input to neurons is $\mathcal{O}(1)$. In this case, we can write

\begin{equation}\label{eq:scaledvariance_homo}
\sigma_{{pq}}^2=\frac{\sigma^2}{N}
\end{equation}
and refer to the parameter $\sigma^2$ as the {\it scaled variance}.
This scaling ensures that, in absence of correlations between synapses, the connectivity matrix exhibits a random spectrum of radius independent of $N$ and given by $\sigma$ \cite{edelman2005random,rajan2006eigenvalue,kadmon2015transition,tao2023topics}. If $\lambda_0<-\sigma$, the mean part of the connectivity leads to a negative outlier \cite{rajan2006eigenvalue,ostojic2014two,mastrogiuseppe2018linking, dahmen2019second}.
In addition to the first-order statistics, we will assume that the synaptic strengths exhibit correlations described by uniform chain-motif statistics $\tau^c$ and reciprocal-motif coefficient $\tau^r$ defined in Eq.\eqref{eq:LocallyDefinedStats}. Numerical procedures for generating the corresponding connectivity matrices are described in Appendix~\ref{ap:networkconstructionGaussian}.

The equation for the outlying eigenvalues in this homogeneous case is derived in Appendix \ref{ap:homogeneousGauss}.
 In Appendix \ref{ap:heterogeneousGauss}, we consider the more general situation where the variances are different in blocks connecting different populations.

\begin{figure*}[ht]
\includegraphics[width=1\textwidth]{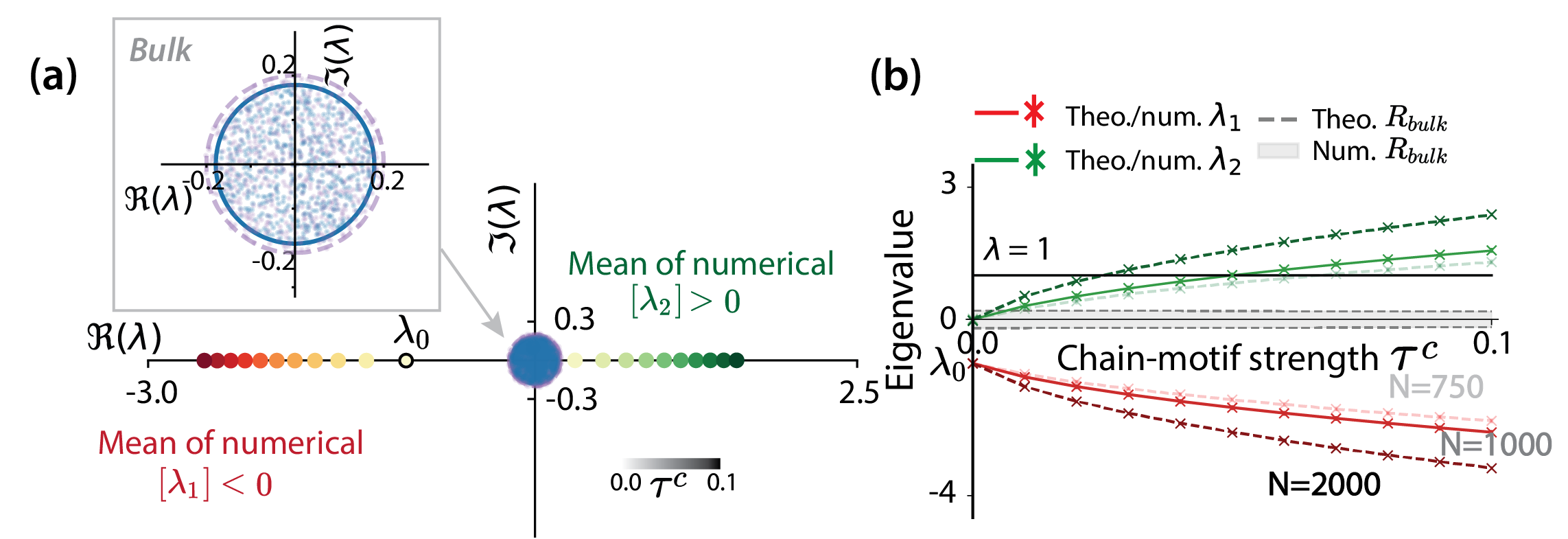}
\caption{\label{fig:F1_ConnStatsOutliers}
Impact of chain motifs on the eigenvalues of the connectivity matrix for fully connected excitatory-inhibitory networks.
(a) Eigenspectrum in the complex plane. The spectrum consists of a circular bulk (magnification in inset), within which the eigenvalues are continuously distributed, and isolated outliers. Inhibition dominated networks with independent synapses give rise to a single negative outlier $\lambda_0$ (black circle). Progressively increasing the strength $\tau^{c}$ of chain motifs (from light to dark), this negative outlier (yellow to red) decreases from the original $\lambda_0$, while an additional positive outlier emerges and increases (light to dark green). The dots show numerically-determined outlying eigenvalues averaged over $30$ networks of $N=1000$ neurons. The inset shows the 
eigenvalue bulk for a single network realization  with $\tau^c=0$ (purple) and $\tau^c=0.1$ (blue).
(b) Dependence of outlying eigenvalues on the strength $\tau^{c}$ of chain motifs and the network size $N$.
Numerically obtained eigenvalue outliers from $30$ network realizations (markers) are shown alongside the theoretical predictions (lines) calculated using Eq.\eqref{eq:twoOutliersMaintext}. The dominant negative outlier is depicted in red, while the emergent positive outlier is denoted in green.   The gray area indicates the radius of the eigenvalue bulk in networks with a size of $N=1000$, with dashed lines indicating the theoretical values \cite{dahmen2020strong}. 
As a control, we fix $\lambda_0$ across networks with different $N$, by scaling the mean synaptic weight $J^0_{pq}$ as $1/N$.
All parameter values are given in TABLE \ref{tab:parameters}.
}
\end{figure*}

\subsubsection{Eigenvalues}

Applying the analytical approach outlined in Sec.\ref{sec:EigenvaluesMethod} to networks featuring homogeneous chain and reciprocal motifs,  in Eq.~\eqref{eq:motifQuadraticZ} the $N(N-1)$ off-diagonal entries in $[\mathbf{Z}^2]$ are non-zero, equal and determined by the chain motif strength $\tau^c$, while the $N$ diagonal entries are determined by the reciprocal  motif strength $\tau^r$. Altogether, $[\mathbf{Z}^2]$ is a sum of a diagonal matrix and a uniform, unit-rank matrix (Eq.~\eqref{eq:ZsquareDecompose_Maintext}); substituting this into Eq.~\eqref{eq:matrixInverse_main} leads to the following equation for eigenvalue outliers:

\begin{equation}\label{eq:polynomial_EigenvalueGaussian}
\begin{aligned}
\lambda^2-\lambda_0\lambda-\Delta^2=0,
\end{aligned}
\end{equation}
where we assumed $\lambda\neq 0$, and
\begin{equation}
\Delta^2=\sigma^2\tau^{c}(N-1)+\sigma^2\tau^{r}.
\end{equation}

This second-order polynomial equation has two solutions

\begin{equation}\label{eq:twoOutliersMaintext}
\begin{aligned}
\lambda_{1,2} &=\frac{\lambda_0\mp\sqrt{\lambda_0^2+4\Delta^2}}{2}.
\end{aligned}
\end{equation}
In the following, we refer to $\Delta$ as the eigenvalue perturbation strength. Since we assumed $\lambda_0\leq 0$, for $\Delta=0$, $\lambda_1=\lambda_0$ and $\lambda_2=0$. As $\Delta$ is increased, $\lambda_1$ decreases from $\lambda_0$ and generates an increasingly negative real outlier, while $\lambda_2$ increases from zero and can give rise to a positive real outlier if it emerges from the bulk of the eigenspectrum.
For networks with only chain motifs we have
\begin{equation}\label{eq:delta_chain}
\Delta^2=\sigma^2\tau^{c}(N-1),
\end{equation}
so that the modulus of the perturbed eigenvalues scales as $\sqrt{\tau^cN}$ (Fig~\ref{fig:F1_ConnStatsOutliers}). In particular, if the network size $N$ is increased at fixed $\tau^c$, the positive eigenvalue emerges from the random bulk and eventually crosses unity.
This implies that positive chain motif strengths fundamentally induce an instability in network dynamics, unless they scale inversely with network size,  i.e. $\tau^c \sim 1/N$.

For networks with only homogeneous reciprocal motifs $\tau^r$, the
eigenvalue perturbation is instead independent of network size and given by
\begin{equation}\label{eq:eigvperturbationStrength_rec_Gauss}
\Delta^2=\sigma^2\tau^{r}.
\end{equation}
Compared to chain motifs, the perturbation term is independent of $N$, so that reciprocal motifs have a much weaker effect on outlying eigenvalues \cite{shao2023relating}. This difference can be traced back to  Eq.~\eqref{eq:ZsquareDecompose_Maintext}, where reciprocal motifs appear only in the $N$ diagonal entries of $[\mathbf{Z}^2]$, while chain motifs determine the $N(N-1)$ off-diagonal elements, which leads to a factor $N-1$ in Eq.\eqref{eq:delta_chain}. 
Convergent and divergent motifs have even weaker effects on outlying eigenvalues (Fig.~\ref{fig:comparable_ReciprocalMotifs}), as they do not appear in $[\mathbf{Z}^2]$, and enter only the averages of higher powers of $\mathbf{Z}$ at the sub-dominant order in $1/N$ (Appendix \ref{ap:GeneralMotifs}).

Altogether, the key result of our analysis is therefore that {\it chain motifs} generate a positive outlying eigenvalue which increases with network size, and therefore potentially induces an instability by pushing the real part of the outliers above 1, even in otherwise stable, inhibition-dominated networks. 

\begin{figure*}[ht]
\includegraphics[width=1.0\textwidth]{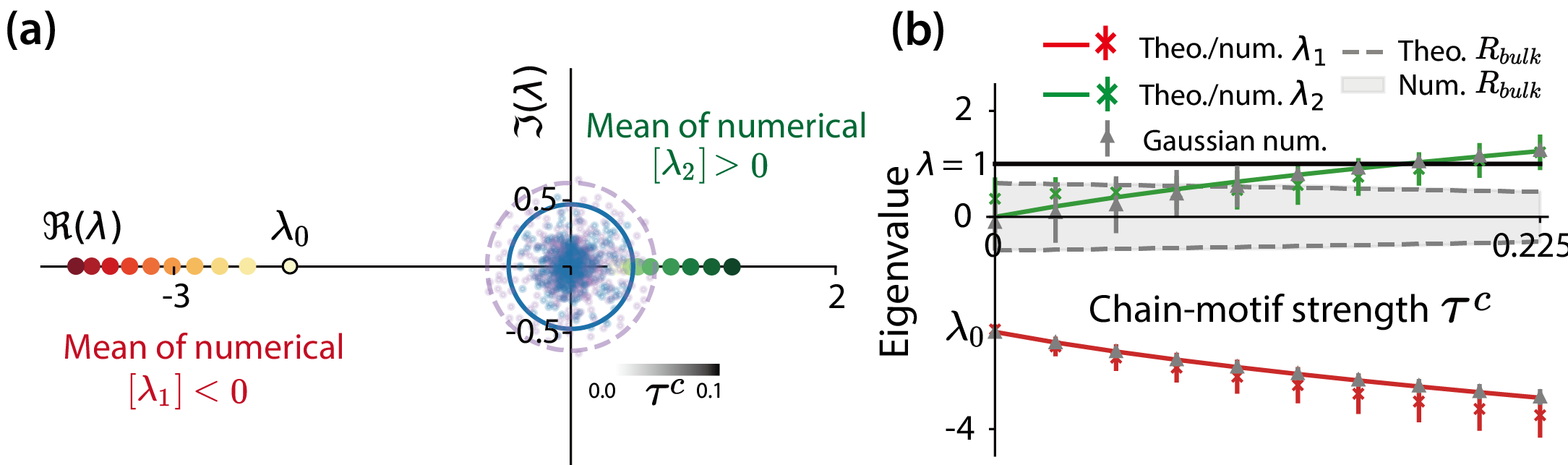}
\caption{\label{fig:EigenvaluesAdjacencyMatrix}
Impact of chain motifs on the eigenvalues of the connectivity matrix for sparsely connected excitatory-inhibitory networks.
(a) Eigenspectrum in the complex plane.  Inhibition dominated networks with independent synapses give rise to a single outlier $\lambda_0$ (black circle). Progressively increasing the strength $\tau^{c}$ of chain motifs (from light to dark), this negative outlier (yellow to red) decreases from the original $\lambda_0$, while an additional positive outlier emerges and increases (light to dark green). The dots show numerically-determined outlying eigenvalues averaged over $30$ networks of $N=1500$ neurons. The eigenvalue bulk for a single network realization  is shown for $\tau^c=0$ (purple) and $\tau^c=0.225$ (blue).
(b) Dependence of outlying eigenvalues on the strength $\tau^{c}$ of chain motifs.
Numerically obtained eigenvalue outliers from $30$ network realizations (markers) are shown alongside the theoretical predictions (lines).
The dominant negative outlier is depicted in red, while the emergent positive outlier is shown in green. The  colored solid lines show the theoretical results calculated using Eqs.~\eqref{eq:twoOutliersMaintext_sparse},~\eqref{eq:chainperturb}. The red and green asterisks with error bars represent the results numerically obtained from actual sparse networks (same as subplot (a)), and the gray triangles with error bars represent the results numerically obtained from the equivalent Gaussian networks. 
The gray area indicates the radius of the eigenvalue bulk in networks with a size of $N=1500$, with dashed lines indicating the theoretical values 
(\cite{dahmen2020strong} for details).
All parameter values are given in TABLE \ref{tab:parameters}.
}
\end{figure*}
\subsection{\label{subsecs:EigenvalueAdj}Sparse excitatory-inhibitory networks}
Next, we turn to sparse excitatory-inhibitory networks and show that chain motifs induce similar outlying eigenvalues to fully-connected networks. We then examine the conditions under which sparsity and excitation-inhibition balance can stabilize the positive outlying eigenvalue. 

\subsubsection{Setup}
In sparse networks, only a fraction of possible synaptic connections are non-zero. To define sparse networks with second-order motifs, we start by specifying the first-order statistics of non-zero connections, then add second-order statistics and finally assign synaptic weights to non-zero connections.

As in standard sparse networks \cite{van1996chaos,ginzburg1994theory,renart2010asynchronous}, the first-order statistics are set by taking any synaptic weight from population $q$ to population $p$ to be non-zero with probability $c_{pq}$ and zero otherwise.
Here we assume homogeneous connectivity probabilities $c_{pq}=c$ across populations. The marginal distribution of non-zero connections is therefore  expressed as
\begin{equation}\label{eq:firstOrderSparse}
\begin{aligned}
Prob(J_{ij}\neq 0)&=c, \\
Prob(J_{ij}=0)&=1-c.
\end{aligned}
\end{equation}

To add chain and reciprocal synaptic motifs, i.e. specific correlations between synapses, we introduce two supplementary parameters $\rho^c$ and $\rho^r$ that define the joint probabilities of non-zero connections across pairs of synapses
\begin{equation}\label{eq:secondOrderSparse}
\begin{aligned}
Prob(J_{ij}\neq 0,J_{jk}\neq 0)&=(1-\delta_{ik})\rho^c+\delta_{ik}\rho^r, \\
Prob(J_{ij}=0,J_{jk}\neq 0)&=(1-\delta_{ik})(c-\rho^c)+\delta_{ik}(c-\rho^r),\\
Prob(J_{ij}\neq 0,J_{jk}=0)&=(1-\delta_{ik})(c-\rho^c)+\delta_{ik}(c-\rho^r),\\
Prob(J_{ij}=0,J_{jk}=0)&=(1-\delta_{ik})(1-c-(c-\rho^c))\\
&+\delta_{ik}(1-c-(c-\rho^r)).
\end{aligned}
\end{equation}
We denote the parameters $\rho^c$ and $\rho^r$ as the {\it occurence probabilities} for respectively chain and reciprocal motifs.
Note that here we implicitly assume a homogeneous motif distribution across different populations. In this framework, independent connections correspond to $\rho^c=c^2$ and $\rho^r=c^2$.

Finally we specify synaptic weights for non-zero synapses. As in the fully connected case,  we focus on networks consisting of two populations, an excitatory and an inhibitory one, with respectively  $N_E=\alpha_E N$ and $N_I=\alpha_I N$ neurons. For simplicity, we assume that all excitatory and all inhibitory
synapses have identical weights, respectively $J$ and  $-gJ$ \cite{brunel2000dynamics},  where $g$ represents the ratio of the amplitudes of excitatory synapses to inhibitory synapses. 

In practice, we apply the SONET algorithm \cite{zhao2011synchronization}, see Appendix \ref{ap:Sparsenetwork}.

\subsubsection{Gaussian approximation \label{sec:gauss_approx}}

Previous studies have found that when the distribution of synaptic couplings satisfies the assumptions of the central limit theorem, the spectral properties of the connectivity matrix are well described by Gaussian connectivity with identical first two moments \cite{rajan2006eigenvalue,kadmon2015transition,shao2023relating,herbert2022impact,dahmen2019second}. 
To determine the outlying eigenvalues in sparse networks with chain motifs, we compute the corresponding first and second moments of the distribution of synaptic weights, and  insert them into  expressions  for the outlying eigenvalues in Gaussian networks.  We then compare these analytic predictions with  numerical estimations of outlying eigenvalues in sparse networks.

The means and variances of the  distributions of synaptic weights for the excitatory and inhibitory populations are given by 
\begin{equation}\label{eq:meanvariance_Gaussianapproximation}
\begin{aligned}
J^0_{EE}=J^0_{EI} = cJ,\,\,\,\sigma_{EE}^2= \sigma_{EI}^2 &=c(1-c)J^2 \\
J^0_{II}=J^0_{IE} = -cgJ,\,\,\,\sigma_{II}^2= \sigma_{IE}^2 &=c(1-c)g^2J^2. \\
\end{aligned}
\end{equation}

Using the definitions in Eq.\eqref{eq:LocallyDefinedStats}, the correlation coefficients of chain motifs between different populations are given by:
\begin{equation}\label{eq:abundance2Covariance}
\begin{aligned}
\tau^c_{EE}&=(\rho^c-c^2)/(c(1-c)),\\
\tau^c_{EI}&=-(\rho^c-c^2)/(c(1-c)),\\
\tau^c_{IE}&=-(\rho^c-c^2)/(c(1-c)), \\
\tau^c_{II}&=(\rho^c-c^{2})/(c(1-c)).
\end{aligned}
\end{equation}

Analogously, 
the correlation coefficients of reciprocal motifs $\tau_{pq}^r,p,q\in\{E,I\}$ are 
\begin{equation}\label{eq:abundance2Covariance_rec}
\begin{aligned}
\tau^r_{EE}&=(\rho^r-c^2)/(c(1-c)),\\
\tau^r_{EI}&=-(\rho^r-c^2)/(c(1-c)),\\
\tau^r_{IE}&=-(\rho^r-c^2)/(c(1-c)), \\
\tau^r_{II}&=(\rho^r-c^{2})/(c(1-c)).
\end{aligned}
\end{equation}

In contrast to fully-connected networks, in sparse networks the mean, variances and correlations of synaptic weights are therefore not independent parameters, but are jointly controlled by the connection probability $c$, the E and I synaptic weights $J$ and $-gJ$ and the motif occurrence probabilities $\rho^c$ and $\rho^r$.
The relative sign of excitation and inhibition in particular necessarily implies differences in correlation coefficients across different blocks of the connectivity matrix. In addition,  
the variances also vary across different blocks of the connectivity matrix, while in Sec.~\ref{subsecs:EigenvalueGauss} we assumed all those parameters were homogeneous.
An over-representation of EE- and II-type motifs with respect to independent connectivity ($\rho^c,\rho^r>c^2$) yields positive correlation coefficients $\tau_{EE}=\tau_{II}:=+\tau^c$, while an over-representation of EI- and IE-type motifs leads to negative correlation coefficients $\tau_{EI}=\tau_{IE}:=-\tau^c$. 

Inserting the expressions for the means in Eqs.~\eqref{eq:gauss_lambda0}, yields an expression of the outlying eigenvalue in the absence of correlations:

\begin{equation}
\begin{aligned}
\lambda_0 &= cN_EJ - cN_IgJ\\
&=(\alpha_E-g\alpha_I)cJN.
\end{aligned}
\end{equation}

Applying the approach outlined in Sec.\ref{sec:EigenvaluesMethod} to networks with heterogeneous variances and correlations leads to expressions for outlying eigenvalues similar to Eq.~\eqref{eq:twoOutliersMaintext} (Appendix \ref{ap:heterogeneousGauss}).

\begin{equation}\label{eq:twoOutliersMaintext_sparse}
\begin{aligned}
\lambda_{1,2}&=\frac{\lambda_0\mp\sqrt{\lambda_0^2+4\Delta^2}}{2},\\
\end{aligned}
\end{equation}
where for chain motifs
\begin{equation}\label{eq:chainperturb}
\Delta=JN(\alpha_E-g\alpha_I)\sqrt{c(1-c)\tau^c}.
\end{equation}

Similarly to fully connected networks, in sparse networks chain motifs therefore lead to the emergence of a positive outlier in inhibition-dominated networks (Fig~\ref{fig:EigenvaluesAdjacencyMatrix}).
Comparing Eq.~\eqref{eq:chainperturb} and Eq.~\eqref{eq:delta_chain} reveals that in sparse networks, the eigenvalue perturbation strength scales with $N$, similarly to fully-connected networks, so that the positive eigenvalue potentially leads to an instability. However in sparse excitatory-inhibitory networks, the perturbation strength also depends on the connection probability strength $c$ and it necessarily contains two opposing terms given by the excitation-inhibition balance $\alpha_E-g\alpha_I$.
In the following, we will examine how different assumptions on the connectivity probability and synaptic weights impact the scaling of outlying eigenvalues with the network size $N$ and the stability of the network.

\begin{figure}[b]
\includegraphics[width=0.48\textwidth]{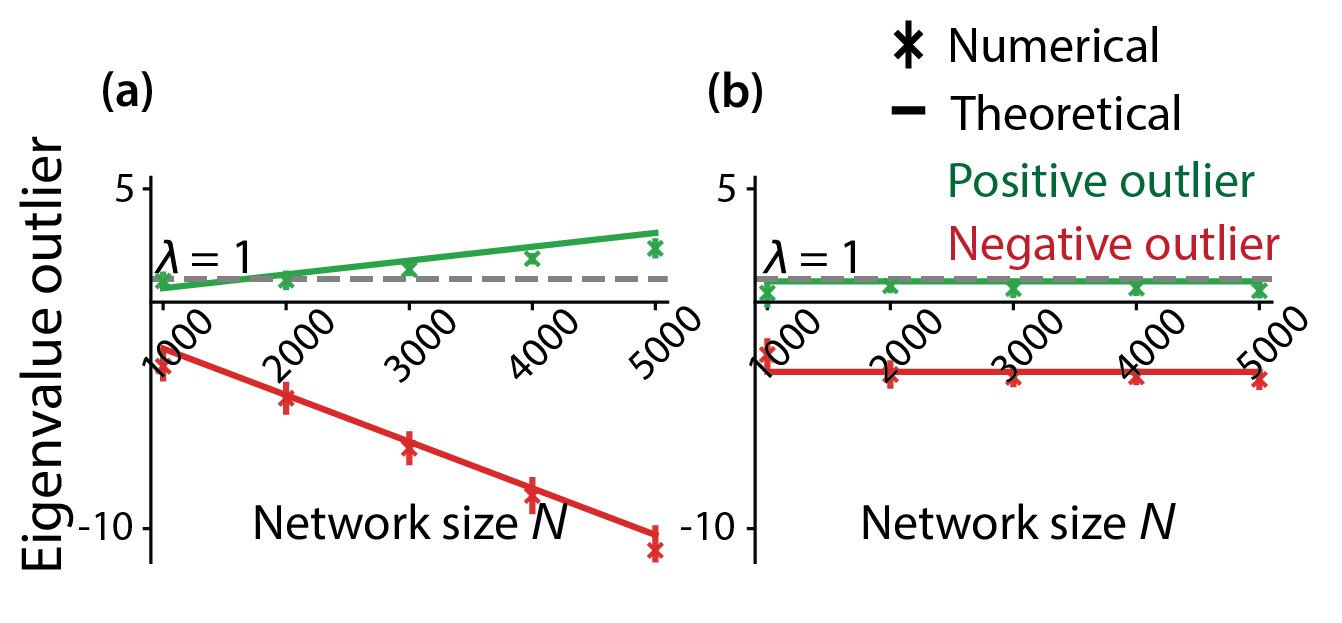}
\caption{\label{fig:networkscaling}
Dependence of outlying eigenvalues on the network size $N$ ranging from 1000 to 5000 for sparse excitatory-inhibitory connectivity in two different scaling limits.
(a) Networks in the strongly connected regime with a constant connectivity probability of $c=0.2$ and a chain motif probability of $\tau^c=0.15$ ($\rho^c=0.064$). 
(b) Networks in the weakly connected regime with a fixed number of connections and  chain motifs, where $C_E=240$ and $k^c_{EE}=92160$.
The asterisks with error bars indicate the mean and standard deviation of numerically obtained eigenvalue outliers from 30 network instances. 
Other parameters: $g=6.8,~\gamma=1/4,~J=0.0129$.
}
\end{figure}

\subsubsection{Strongly connected regime}
We consider first a scaling limit in which the connection probability $c$ and chain motif statistic $\rho^c$ (and therefore $\tau^c$) are independent of $N$, and  the synaptic weight $J$ is of order $1/\sqrt{N}$. This scaling is analogous to the strongly connected regime in networks without motifs \cite{renart2010asynchronous}.
The perturbation term $\Delta$ in Eq.~\eqref{eq:chainperturb} then scales as $\sqrt{N}(\alpha_E-g\alpha_I)$. Consequently, the influence of chain motifs on the eigenvalue outliers amplifies with the network size (see Fig~\ref{fig:networkscaling}(a)) and leads to an instability, similarly to our  findings in Gaussian networks.

The growth of the perturbation term $\Delta$ can only be limited if the balance factor $\alpha_E-g\alpha_I$ scales as $1/\sqrt{N}$.
In the strongly connected regime, ensuring network stability thus requires  tight balance, with $\alpha_E-g\alpha_I$ approaching zero \cite{hennequin2017inhibitory}. 
This balance requirement also ensures that  $\lambda_0$, and the negative outlier, take values independent of network size.

\subsubsection{Weakly connected regime}
We next consider a  second type of scaling with network size, where the connection probability $c$ and chain motif occurrence
$\rho^c$ decrease with $N$ in such a way that
the number $C_p$ of incoming synaptic connections for each neuron, the number of motifs $k^c_{pq}$ and the synaptic strengths $J$ are fixed independently of $N$. This scaling is analogous to the weakly connected regime in networks without motifs \cite{brunel2000dynamics}.
Here, $C_p$ represents the number of excitatory (if $p=E$) or inhibitory (if $p=I$) synapses targeting the same neuron, while $k^c_{pq}$ denotes the number of chain motifs originating from neurons in population $q$, passing through neurons in population $p$ and ultimately targeting an individual neuron. These parameters are related to the connection probability $c$ and motif occurrence probability $\rho^c$ via 
\begin{equation}\label{eq:relation_firstseconddesigns}
C_p = cN_p,\,\,\,k_{pq}^c = \rho^cN_pN_q,\,\,\,p,q\in\{E,I\}.
\end{equation}
The weakly connected regime therefore implies that the connection probability scales as $1/N$ and motif occurrence probability as $1/N^2$.

Expressing the eigenvalue perturbation term in Eq.~\eqref{eq:chainperturb} as a function of these parameters, we get
\begin{equation}\label{eq:chainperturb_fixednum}
\begin{aligned}
\Delta=J(1-g\frac{\alpha_I}{\alpha_E})\sqrt{(k^c_{EE}-C_E^2)}.
\end{aligned}
\end{equation}
In the weakly connected regime, the eigenvalue outliers are therefore independent of  the network size $N$(Fig~\ref{fig:networkscaling}(b)). The stability of the network in the presence of chain motifs is therefore ensured without assuming tight balance between excitation and inhibition.  

\section{\label{secs:EigenvectorDynamics}Population-averaged responses to uniform inputs}

In this section, we examine how chain motifs affect steady-state responses to external inputs. We focus on population-averaged responses to inputs that are uniform over each population, and specifically examine under which conditions paradoxical responses emerge. To this end, we compare two methods: (i) predicting responses from a low-rank approximation of the recurrent connectivity matrix; (ii) predicting responses from an effective deterministic matrix $\mathbf{J}^{eff}$.
We start by outlining the general approach, and then apply it to fully-connected and sparse excitatory-inhibitory networks. 

\subsection{Low-rank approximation}

Following the approach introduced in Sec.~\ref{secs:LowrankApproach}, we form a low-rank approximation of the connectivity matrix $\mathbf{J}$ by truncating its eigen-decomposition to keep only modes corresponding to the outlying eigenvalues determined in the previous section.  We obtain  the corresponding left and right connectivity vectors  numerically. Within this low-rank approximation, the steady-state  response functions of the network are given by Eqs.~\eqref{eq:steady_state_low_rank} and \eqref{eq:responseLowrank}.

We  focus on changes in the average activity of different populations in response to  inputs that can differ among populations, but are uniform within each population, so that ${I}^{ext}_j = {I}^{ext}_q$ for all $j$ in population $q$.
Averaging Eq.~\eqref{eq:responseLowrank} over the connectivity distribution, the mean response $\chi_{pq}$ of a neuron $i$ in population $p$ to a uniform input to neurons in population $q$ can be expressed as

\begin{eqnarray}
\chi_{pq} &:=&
\left[\sum_{j\in N_q}\chi_{ij}\right]  \\
&=& \delta_{pq}+\alpha_q \sum_{r=1}^R\frac{a_{m_r}^p a_{n_r}^q}{1-\lambda_r}. \label{eq:PopresponseLowrank}
\end{eqnarray}
In the large network limit ($N_p\to \infty$), where $[\cdot]$ stands for averaging over connectivity.
For any  vector $\boldsymbol{x}\in\mathbb{R}^{N}$, $a_{x}^p$ is the mean value of its entries within population $p$:

\begin{equation}
a_{x}^p = [x_i]\,\,\, \mathrm{for }\,\, i\,\,  \mathrm{ in \,\,population  }\,\, p.
\end{equation}
When averaging Eq.~\eqref{eq:responseLowrank}, we assumed that $m_i^{(r)}$ and $n_j^{(r)}$ are uncorrelated for $i\neq j$ so that $\left[m_i^{(r)}n_j^{(r)}\right]=\left[m_i^{(r)}\right]\left[n_j^{(r)}\right]=a_{m_r}^p a_{n_r}^q$.

Eq.~\eqref{eq:PopresponseLowrank} shows that the mean values $a_{m_r}^p$ and $a_{n_r}^q$ are the only statistics of connectivity vectors $\boldsymbol{m}^{(r)}$ and $\boldsymbol{n}^{(r)}$  that affect population-averaged responses to uniform inputs. We determine these values numerically by averaging entries of numerically-obtained eigenvectors.

\begin{figure*}[ht]
\includegraphics[width=0.98\textwidth]{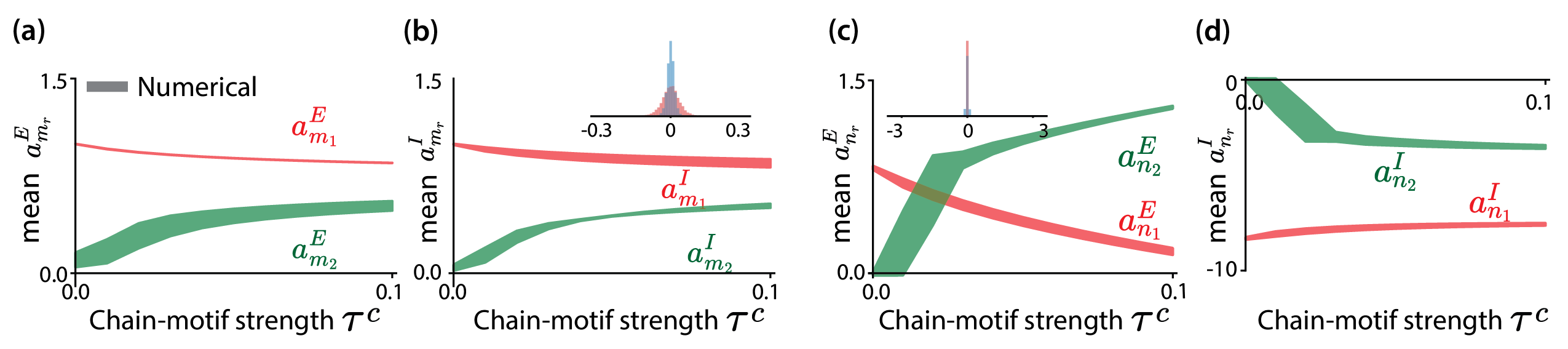}
\caption{\label{fig:F1_ConnStatsVectors}
Impact of chain motifs on population-averaged mean of entries on connectivity vectors for Gaussian networks.
(a, b) Population-averaged mean values of entries $m_i^{(r)p}$ on the right connectivity vectors. Subplots (a) and (b) respectively show the mean values for $p=E$ and  $p=I$. 
(c, d) Population-averaged mean values of entries $n_i^{(r)p}$ on the left connectivity vectors. Subplots (c) and (d) respectively show the mean values for $p=E$ and  $p=I$. 
The insets show the distribution of the mean values of the elements in the connectivity vectors corresponding to the eigenvalues in the bulk (red: excitatory population; blue: inhibitory population).
The additional eigenvalue emerges from the bulk when $\tau^c=0.011$.
Other network parameters see TABLE~\ref{tab:parameters}.}
\end{figure*}

\subsection{Effective connectivity approximation}
An alternative approach is to directly average the response function in Eq.~\eqref{eq:response_full_rank} over realizations of the random connectivity \citep{hu2018feedback}. We show in Appendix \ref{ap:EffectiveConnectivityMatrix} that, at leading order in $N$, the average response is equivalent to the response generated by the effective connectivity matrix
\begin{equation} \label{eq:Jeff_def}
\mathbf{J}^{eff} = \mathbf{J}^0+\left[\mathbf{Z}^2\right].
\end{equation}
This effective connectivity is a deterministic matrix consisting in general of $P\times P$ blocks. Within each block, all entries are equal, and given by a combination of the mean synaptic strength between two populations and the strength of chain and reciprocal motifs that determine the entries of $\left[\mathbf{Z}^2\right]$ (Eq.~\eqref{eq:motifQuadraticZ}).
$\mathbf{J}^{eff}$ is therefore  a low-rank matrix, of rank at most $P^2$, and equivalent to a $P\times P$ matrix. The response predicted by $\mathbf{J}^{eff}$ can then be analyzed using methods developed for deterministic connectivity matrices \citep{tsodyks1997paradoxical,miller2020generalized}.

\subsection{\label{subsec:EigenvectorGauss}Fully connected Gaussian excitatory-inhibitory networks}
\begin{figure}[hbt]
\includegraphics[width=0.50\textwidth]{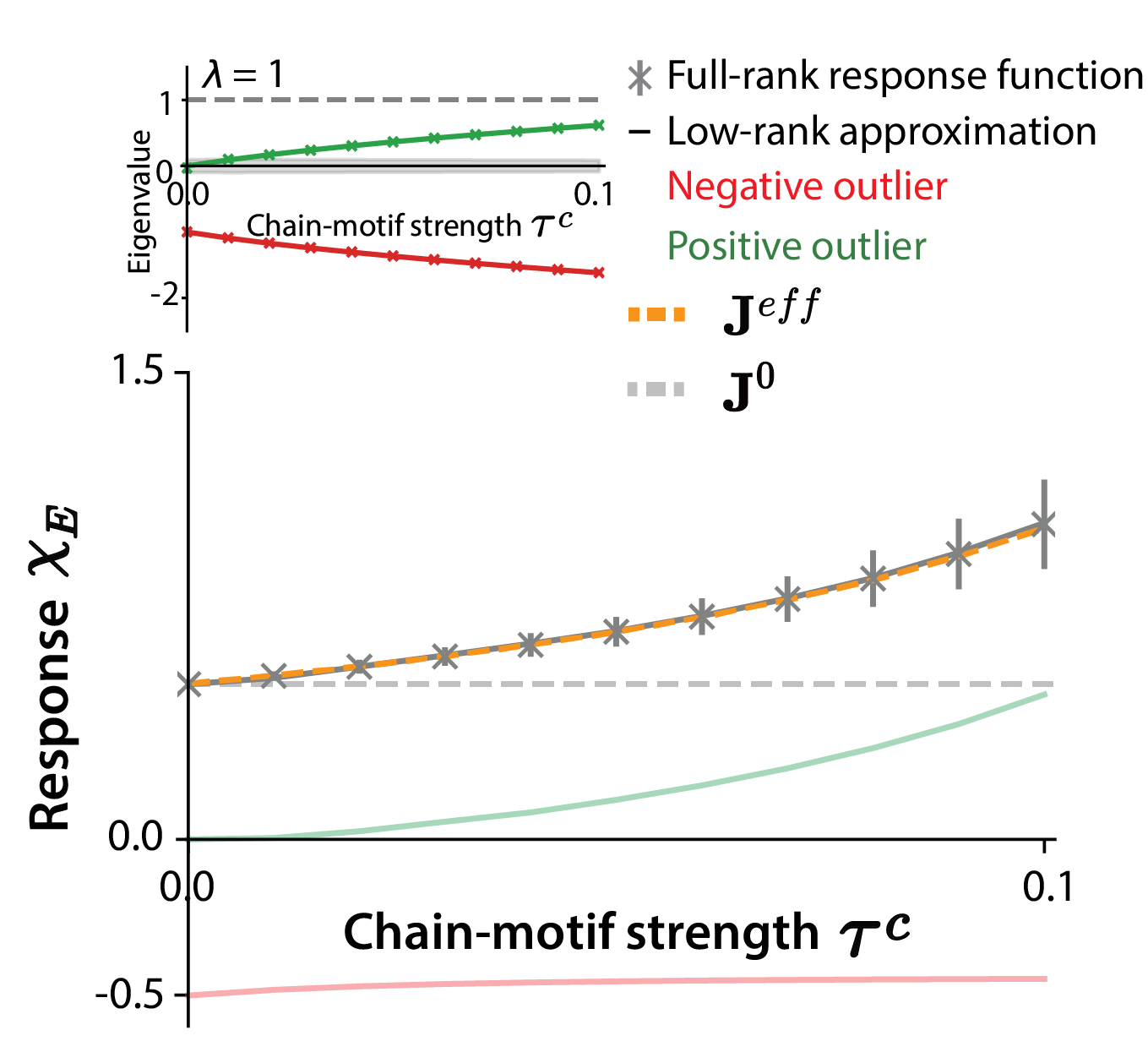}
\caption{\label{fig:dynamicsExternalInp}
Impact of chain motifs on the response to uniform external inputs in fully-connected Gaussian excitatory-inhibitory networks:
 mean response $\chi_{E}$ of the excitatory population as a function of the strength $\tau^{c}$ of chain motifs, which is homogeneous across all blocks of the connectivity matrix. The asterisks with error bars show the mean and s.d. of population-averaged responses in 30 realizations of the full connectivity (Eq.~\eqref{eq:response_full_rank}). 
 The lines represent the predictions of the low-rank approximation in  Eq.~\eqref{eq:avgFiringrate_deterministicInp}, where the $a_{x}^p$ were obtained numerically by diagonalizing the full connectivity matrix. The gray dashed line is the prediction obtained from the mean connectivity $\mathbf{J}^0$ (Eq.~\eqref{eq:response_mean_connec}), the orange dashed line is the prediction obtained from the effective connectivity $\mathbf{J}^{eff}$. 
The colored light lines show the respective contributions of the two unit-rank terms in the low-rank approximation of the response (Eq.~\eqref{eq:avgFiringrate_deterministicInp}).
Red and green denote respectively the modes corresponding to the negative and positive outliers. Inset: outlying eigenvalues as a function of $\tau^{c}$, compared to the bulk of the radius.  The positive outlier emerges from the bulk for $\tau^{c}=0.011$. All network parameters are given in TABLE~\ref{tab:parameters}.
}
\end{figure}

We start by examining the effect of chain motifs on responses in fully-connected, Gaussian excitatory-inhibitory networks defined in Sec.~\ref{subsecs:EigenvalueGauss}.

In inhibition-dominated networks, we find that the eigenspectrum contains either one or two outliers depending on the strength of chain motifs $\tau^{c}$ (Fig~\ref{fig:F1_ConnStatsOutliers}). The resulting low-rank approximation is therefore of rank one or two.  We refer to the connectivity vectors corresponding  to the negative outlier as 
$\boldsymbol{m}^{(1)}$ and $\boldsymbol{n}^{(1)}$. When the connectivity motifs induce a second outlier, we refer to the additional connectivity vectors as 
$\boldsymbol{m}^{(2)}$ and $\boldsymbol{n}^{(2)}$.

Fig~\ref{fig:F1_ConnStatsVectors} shows the mean values of the entries of the connectivity vectors onto excitatory and inhibitory neurons. Increasing $\tau^{c}$ has a weak effect on the mean values of $\boldsymbol{m}^{(1)}$ (Fig~\ref{fig:F1_ConnStatsVectors} (a),~(b)), which remain close to their unperturbed values, but a stronger effect on the mean of $\boldsymbol{n}^{(1)}$ within the excitatory population (Fig~\ref{fig:F1_ConnStatsVectors} (c)). Remarkably, chain motifs induce comparable mean values for the connectivity vectors  $\boldsymbol{m}^{(2)}$ and $\boldsymbol{n}^{(2)}$ corresponding to the second outlier. In contrast, eigenvectors corresponding to eigenvalues in the bulk of the eigenspectrum have zero mean (insets in Fig~\ref{fig:F1_ConnStatsVectors}).

\subsubsection{\label{subsub:dynsGauss}Response to uniform inputs}
\begin{figure}[htb]
\includegraphics[width=0.50\textwidth]{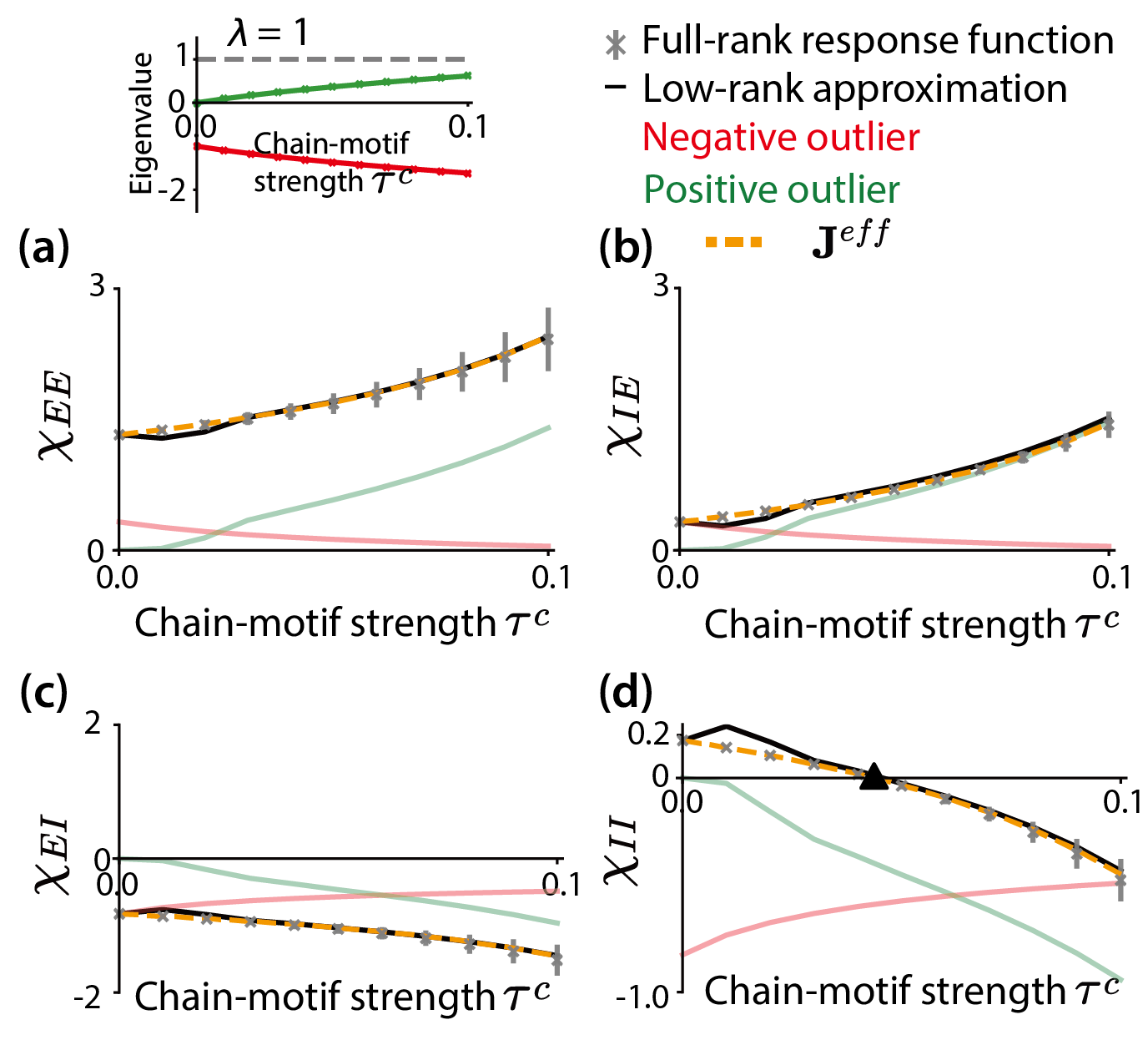}
\caption{\label{fig:GaussianParadoxical_nonISN}
Impact of chain motifs on the mean response $\chi_{pq}$ of population $p$ to uniform inputs to population $q$, in fully-connected Gaussian excitatory-inhibitory networks. (a)  $\chi_{EE}$; (b)  $\chi_{IE}$; (c)  $\chi_{EI}$; (d)  $\chi_{II}$.
The asterisks with error bars show the mean and s.d. of population-averaged responses in $30$ realizations of the full connectivity (Eq.~\eqref{eq:response_full_rank}). 
The orange dashed lines represent the predictions of the effective connectivity $\mathbf{J}^{eff}$, the black triangle in (d) is the prediction of the transition point using the stability of the E-E subnetwork.
The solid lines represent the predictions of the low-rank approximation (Eq.~\eqref{eq:PopresponseLowrank}), with colored lines indicating the contributions from the two unit rank terms (red: negative outlier; green: positive outlier).
In Eq.~\eqref{eq:PopresponseLowrank}, $a_{x}^p$ were obtained numerically by diagonalizing the full connectivity matrix.
The strength $\tau^{c}$ of chain motifs is homogeneous across all blocks of the connectivity matrix.
The mean connectivity parameters  were chosen such that $\chi_{II}>0$ for $\tau^{c}=0$, i.e. the network is in the non-paradoxical regime in absence of inputs. All network parameters are given in TABLE~\ref{tab:parameters}.
}
\end{figure}

We first analyze the impact of chain motifs on 
 the average response $\chi_{p}$ of neurons in population $p$ to a uniform input to all neurons in the network.

From Eq.~\eqref{eq:PopresponseLowrank}, the low-rank approximation for $\chi_{p}$ is given by
\begin{equation}\label{eq:avgFiringrate_deterministicInp}
\begin{aligned}
\chi_{p} = 1+\sum_{r=1}^{R}\frac{a_{m_r}^p}{1-\lambda_r}\left (\sum_{q\in \{E,I\}}\alpha_qa_{n_r}^q\right).
\end{aligned}
\end{equation}

To assess the accuracy of this approximation, we compare it with the  steady-state response obtained using the full connectivity matrix $\mathbf{J}$ (Eq.~\ref{eq:response_full_rank}). We contrast these values with a prediction based solely on the mean connectivity $\mathbf{J^0}=\boldsymbol{m}_0\boldsymbol{n}_0^{\intercal}/N$, for which $a_{m_0}^E=a_{m_0}^I=1$, $a_{n_0}^E=NJ^0$ and $a_{n_0}^I=-NgJ^0$ (Eq.~\eqref{eq:meanstructures_Gaussian}), so that 

\begin{equation}\label{eq:response_mean_connec}
\chi_{E} = \chi_{I} = \frac{1}{1-\lambda_0}
\end{equation}
with $\lambda_0=(\alpha_E-g\alpha_I)J^0N$ (Eq.~\eqref{eq:gauss_lambda0}).

While the mean connectivity $\mathbf{J^0}$ predicts accurately the mean response of random networks without correlations between synapses,
we find that including chain motifs strongly amplifies  the
response of both excitatory and inhibitory populations (Fig~\ref{fig:dynamicsExternalInp}). This amplification is well captured both by the low-rank approximation and the effective connectivity $\mathbf{J}^{eff}$. The accuracy of the low-rank approximation can be explained by the fact that the eigenmodes which correspond to the bulk eigenvalues that were not included in the low-rank description have zero mean components, and therefore do not contribute to the population-averaged response (see insets in Fig~\ref{fig:F1_ConnStatsVectors} (b,c)).

Within the low-rank approximation, we can
distinguish the contributions to the response of the two unit-rank terms (Fig~\ref{fig:dynamicsExternalInp} colored lines). This decomposition reveals that the  unit-rank term  associated with the negative outlying eigenvalue has minimal influence on the amplification induced by chain motifs. Conversely, the amplification is due to the second term that corresponds to the newly emerging eigenmode with a positive outlier.
 The amplification is therefore specific to the outlier induced by chain motifs.
In contrast, networks featuring only reciprocal motifs do not give rise to this outlier, and therefore, there is no noticeable amplification as $\tau^r$ increases (Fig~\ref{fig:ReciprocalDyns}(a,~b)).

\subsubsection{\label{subsub:paradoxicalGauss}Paradoxical responses}

We next use the low-rank approximation to examine the mean responses $\chi_{pq}$ of population $p$ to uniform inputs to population $q$.
Fig~\ref{fig:GaussianParadoxical_nonISN} shows that  chain motifs strongly modulate all components of this response matrix. 
Focusing on the diagonal entries we denote the response of population $p$ as paradoxical if $\chi_{pp}<0$, i.e. if it responds in a direction opposite to the input it receives. Fig~\ref{fig:GaussianParadoxical_nonISN} (d) shows that increasing the strength of chain motifs changes the sign of $\chi_{II}$ from positive to negative, thereby switching the response of inhibitory neurons from non-paradoxical in the absence of chain motifs to  paradoxical as $\tau^{c}$ is increased.

In networks without correlations between synapses, previous works have shown that the sign of the response of the inhibitory population is controlled by the strength of the excitatory feedback \cite{tsodyks1997paradoxical,rubin2015stabilized,litwin2016inhibitory}.
Within our framework, the responses in an uncorrelated network are well predicted by using the mean connectivity $\mathbf{J^0}=\boldsymbol{m}_0\boldsymbol{n}_0^T/N$ in Eq.~\eqref{eq:PopresponseLowrank}, which leads to:

\begin{equation}\label{eq:nonISN_GaussMean}
\begin{aligned}
\chi_{II}&= \frac{1-N_EJ^0}{1-\lambda_0}.
\end{aligned}
\end{equation}
A paradoxical response of the inhibitory population corresponds to  $N_EJ^0>1$, in agreement with previous results \cite{tsodyks1997paradoxical,rubin2015stabilized,litwin2016inhibitory}.
As  the chain motif strength $\tau^c$ is gradually increased, the change in the response is dominated by the second term in the low-rank approximation in Eq.~~\eqref{eq:PopresponseLowrank}:

\begin{equation}\label{eq:IIParadoxical_chn_Gauss}
\begin{aligned}
\chi_{II}&=1+\frac{\alpha_I a_{m_1}^Ia_{n_1}^I}{1-\lambda_1}+\frac{\alpha_I a_{m_2}^Ia_{n_2}^I}{1-\lambda_2}.
\end{aligned}
\end{equation}
For the inhibitory population, the eigenvector statistics (Fig~\ref{fig:F1_ConnStatsVectors} (d)) show that this second term is negative because $a_{n_2}^I<0$ and $a_{m_2}^I>0$. Starting from a non-paradoxical regime with $N_EJ^0<1$ and $\chi_{II}>0$,
as
$\tau^c$ is increased,  
 this negative term progressively reduces the value of the response function and eventually leads to a change in the sign of $\chi_{II}$(Fig~\ref{fig:GaussianParadoxical_nonISN} (d)).

 For parameter values used in Fig~\ref{fig:GaussianParadoxical_nonISN}, this  change of sign is specific to the response $\chi_{II}$ of the inhibitory population. However, because of the large mean values of the unit-rank term corresponding to the positive outlier, $\tau^c$ strongly modulates all components of the response.

 Our results therefore show that the strength of chain motifs can control whether the network is in a paradoxical regime or not.

We next examine the paradoxical responses  from the perspective of effective connectivity, which in this case has a unit-rank structure, and is given by
\begin{equation}
    J^{eff}_{ij} =\left \{
    \begin{matrix}
        J^0+\sigma^2\tau^c,&j\in N_E\\
        -gJ^0+\sigma^2\tau^c,&j\in N_I.\
        \end{matrix}
        \right.
\end{equation}
Using this effective connectivity and Eq.~\eqref{eq:response_full_rank}, we  accurately predict the response function $\chi_{p},~\chi_{pq}$ of the full network  (orange dashed lines in Figs.~\ref{fig:dynamicsExternalInp} and \ref{fig:GaussianParadoxical_nonISN}).

Importantly, for an equivalent deterministic network defined by $\mathbf{J}^{eff}$, the paradoxical response of the inhibitory neuron population is determined by the stability of the E-E subnetwork $\mathbf{J}^{eff}_{EE}$ \cite{miller2020generalized}, which is made up only of synapses from excitatory to excitatory neurons. The only non-zero eigenvalue of this subnetwork is
 $\lambda_{EE}^{eff} = N_E(J^0+\sigma^2\tau^c)$. The inhibitory population will respond paradoxically if this eigenvalue is greater than 1 and the E-E subnetwork becomes unstable \cite{miller2020generalized}. Therefore,  for $N_EJ^0<1$, the eigenvalue of the E-E subnetwork crosses 1 for $\tau^c=(\frac{1}{N_E}-J^0)/\sigma^2$, and this criterion predicts well the appearance of paradoxical inhibitory responses in the full network (Fig.~\ref{fig:GaussianParadoxical_nonISN}, black triangle).

\subsection{\label{sub:EigenvectorSparse}Sparse excitatory-inhibitory networks} 

\begin{figure*}[tb]
\includegraphics[width=0.98\textwidth]{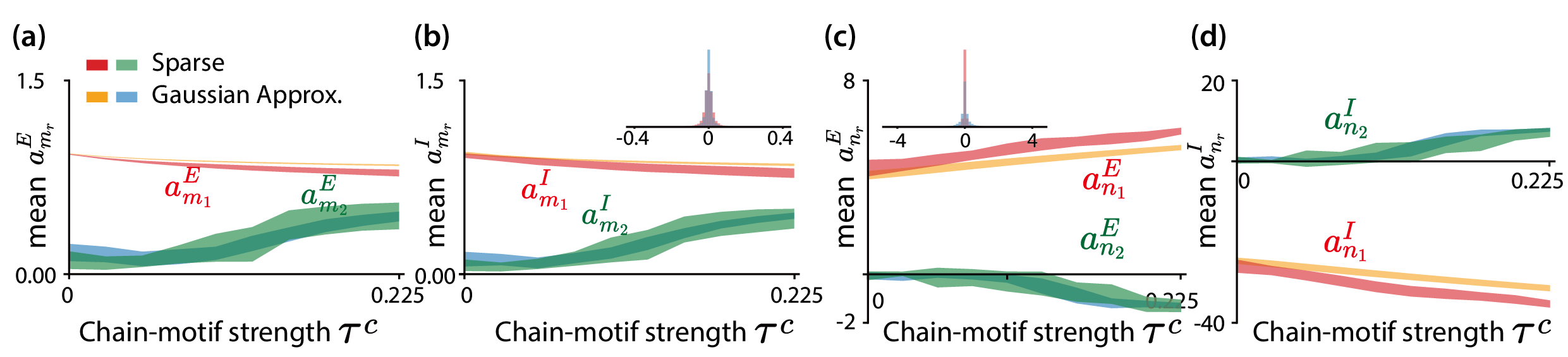}
\caption{\label{fig:AdjacencymatrixVectors}
Impact of chain motifs on population-averaged mean of entries on connectivity vectors for sparse networks (red and green) and their Gaussian approximations (orange and blue).
(a, b) Population-averaged mean values of entries $m_i^{(r)p}$ on the right connectivity vectors. Subplots (a) and (b) respectively show the mean values for $p=E$ and  $p=I$. 
(c, d) Population-averaged mean values of entries $n_i^{(r)p}$ on the left connectivity vectors. Subplots (c) and (d)  respectively show the mean values for $p=E$ and  $p=I$. 
The insets show the distribution of the mean values of the elements in the connectivity vectors corresponding to the eigenvalues in the bulk (red: excitatory population; blue: inhibitory population).
An additional eigenvalue emerges from the bulk  $\tau^c_{EE}=0.11$.
For other network parameters see TABLE~\ref{tab:parameters}.}
\end{figure*}

\begin{figure}
\includegraphics[width=0.50\textwidth]{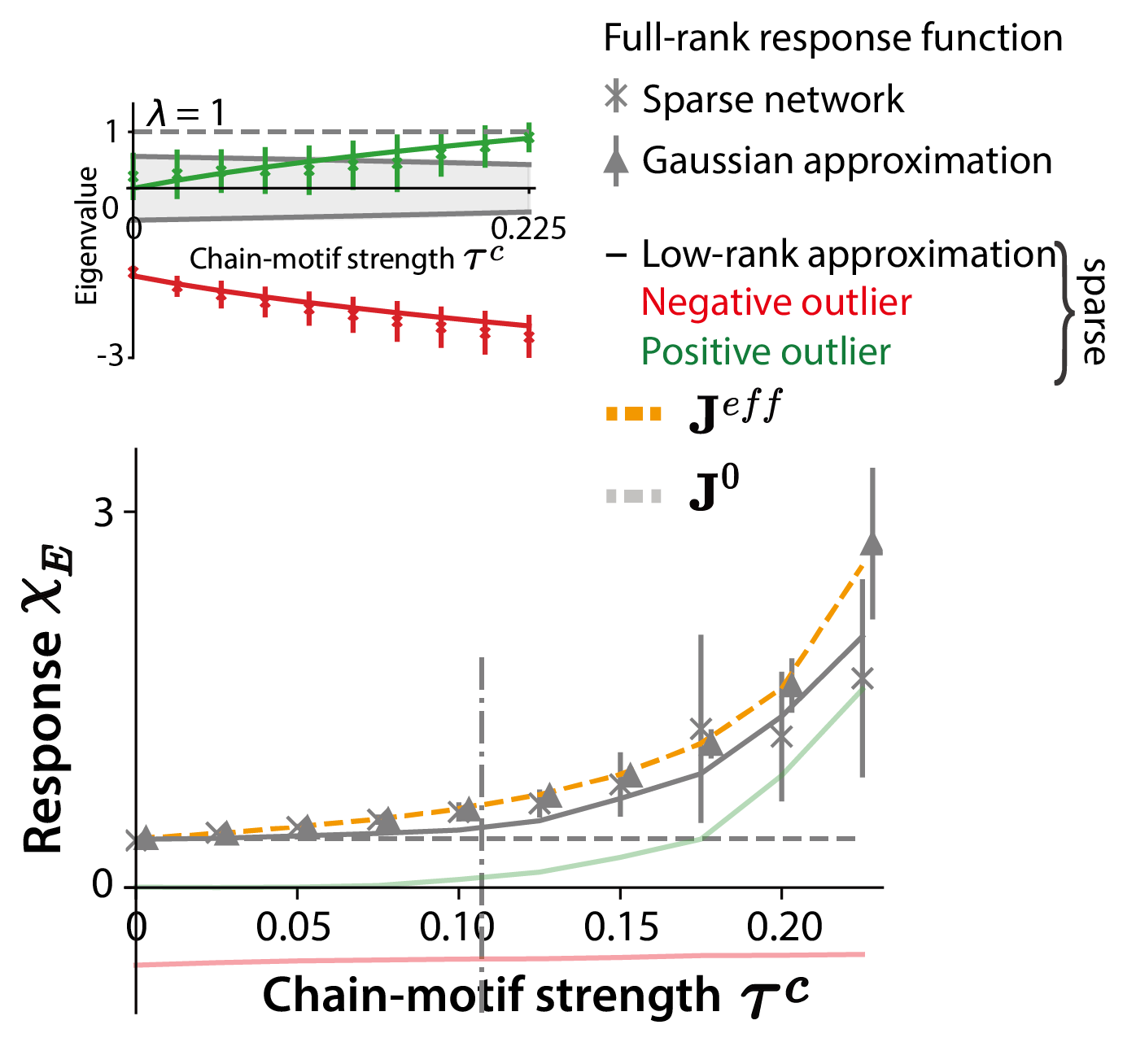}
\caption{\label{fig:dynamicsExternalInpAdjacency}
Impact of chain motifs on the mean response $\chi_{E}$ of excitatory neurons to uniform external inputs in sparse networks, as a function of strength of chain motifs, which is heterogeneous across blocks of the connectivity matrix, with $\tau^c=\tau_{EE}^c=\tau_{II}^c=-\tau_{EI}^c=-\tau_{IE}^c$. The asterisks with error bars show the mean and s.d. of $\chi_{E}$ in 30 realizations of the full sparse network. The triangles with error bars show the mean and s.d. of $\chi_E$ in 30 realizations of the equivalent Gaussian approximation.
The lines represent the predictions of the low-rank approximation in Eq.~\eqref{eq:low-rank-approx}, where the $a_x^p$ were obtained numerically by diagonalizing the sparse connectivity matrix.
The gray dashed line is the prediction obtained from the mean connectivity $\mathbf{J}^0$ (Eq.~\eqref{eq:response_mean_connec}), the orange dashed line is the prediction obtained from the effective connectivity $\mathbf{J}^{eff}$. 
The colored light lines illustrate the respective contributions of the two unit-rank terms in the low-rank approximation of the response of the sparse networks. 
Red and green denote respectively the modes corresponding to the negative and positive outliers.
Inset: outlying eigenvalues as a function of $\tau^c$, compared to the bulk of the radius. The positive outlier emerges from the bulk for $\tau^c=0.11$ (vertical dash-dot line).
For other network parameters, see TABLE~\ref{tab:parameters}.
}
\end{figure}

\begin{figure}
\includegraphics[width=0.50\textwidth]{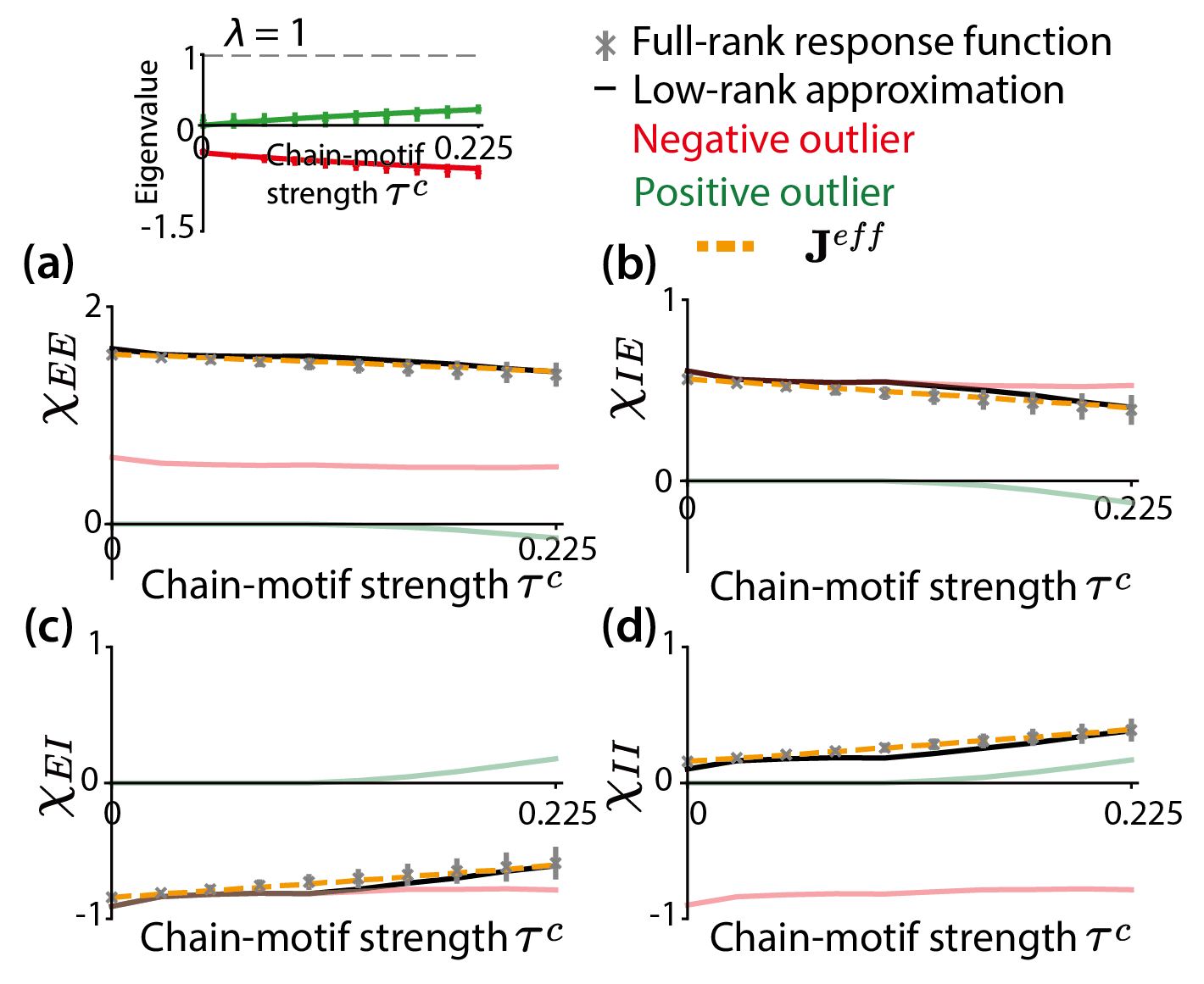}
\caption{\label{fig:SparsenonISN_InhInhInp}
Impact of chain motifs on the mean response $\chi_{pq}$ of population $p$ to uniform inputs to population $q$, in sparse excitatory-inhibitory networks. The subplots show the same quantities as in Fig~\ref{fig:GaussianParadoxical_nonISN}.
The mean connectivity parameters  were chosen such that $\chi_{II}>0$ for $\tau^{c}=0$. All network parameters are given in TABLE~\ref{tab:parameters}.
}
\end{figure}

We found that chain motifs have a major impact on responses in fully connected networks. We next investigated to which extent these results extend to sparse networks, where only a fraction of possible synaptic connections are non-zero, and the strength of chain motifs are controlled by the occurrence probability $\rho^c$ (Eq.~\ref{eq:secondOrderSparse}) which then induces different  values of the correlation coefficients $\tau^c_{pq}$ in different blocks of the connectivity matrix (Eq.~\ref{eq:meanvariance_Gaussianapproximation}).  More specifically, sparse networks with a uniform occurrence $\rho^c$ of chain motifs examined in Sec.~\ref{subsecs:EigenvalueAdj} lead to correlation coefficients $\tau^c_{pq}$ that  obey  $\tau_{EE}^c=\tau_{II}^c=-\tau_{EI}^c=-\tau_{IE}^c$ (Eq.~\ref{eq:meanvariance_Gaussianapproximation}). 

As for fully-connected networks, we approximate the sparse connectivity matrix with a low-rank matrix obtained by keeping only the eigenmodes corresponding to the outlying eigenvalues determined in Sec.~\ref{subsecs:EigenvalueAdj}. We compute the corresponding eigenvectors by numerically diagonizing the sparse connectivity matrix, and then compare their statistics (Fig.~\ref{fig:AdjacencymatrixVectors}) with predictions of perturbation theory for Gaussian networks with identical first and second-order connectivity statistics (Sec.~\ref{sec:gauss_approx},Fig~\ref{fig:AdjacencymatrixVectors}). 
Note that these eigenvectors are in general not sparse, so that the low-rank approximation of a sparse connectivity matrix is in general not sparse \cite{herbert2022impact}.

Similarly to fully-connected networks,
increasing the overall strength of chain motifs by increasing $\rho^c$ amplifies the response $\chi_{p}$ to uniform inputs across the network (Fig~\ref{fig:dynamicsExternalInpAdjacency}). This effect is well captured both by the effective connectivity and by a rank-two approximation, which shows that the amplification is also here due to the positive outlier emerging from the bulk of the eigenspectrum (Fig~\ref{fig:dynamicsExternalInpAdjacency} colored lines).

Inspecting the mean responses $\chi_{pq}$ of population $p$ to uniform inputs in population $q$ however revealed important differences between sparse networks (Figs.~\ref{fig:SparsenonISN_InhInhInp},\ref{fig:SparseISN_2_NonISN})  and fully connected networks  (Fig~\ref{fig:GaussianParadoxical_nonISN}). Indeed, in sparse networks, the mean response of individual populations appeared to change in a direction opposite to fully connected networks when the strength of chain motifs is increased by increasing $\rho^c$. In particular, the mean response of inhibitory neurons to its inputs increases with $\tau^c :=\tau_{EE}^c=\tau_{II}^c=-\tau_{EI}^c=-\tau_{IE}^c$ (Fig~\ref{fig:SparsenonISN_InhInhInp}(d)), while in fully-connected networks it decreases (Fig~\ref{fig:GaussianParadoxical_nonISN}(d)). 
In sparse networks chain motifs can therefore induce a transition from paradoxical inhibitory responses to non-paradoxical responses ($\chi_{II}$ in Fig.~\ref{fig:SparseISN_2_NonISN}), while in fully-connected networks the situation is the opposite.
This difference can be traced back to the mean values $a_{m_r}^p$ and $a_{n_r}^p$ of the entries of connectivity vectors that determine the responses $\chi_{pq}$ through Eq.~\ref{eq:PopresponseLowrank}.
Specifically, the mean values $a_{n_2}^E$ and $a_{n_2}^E$ of the second connectivity vector onto the excitatory and inhibitory populations have opposite signs in sparse networks (Figs~\ref{fig:Supp_AdjacencymatrixVectors},~\ref{fig:AdjacencymatrixVectors} (c,d)) compared to fully-connected networks (Fig~\ref{fig:F1_ConnStatsVectors} (c,d)).

\begin{figure*}[htb]
\includegraphics[width=0.92\textwidth]{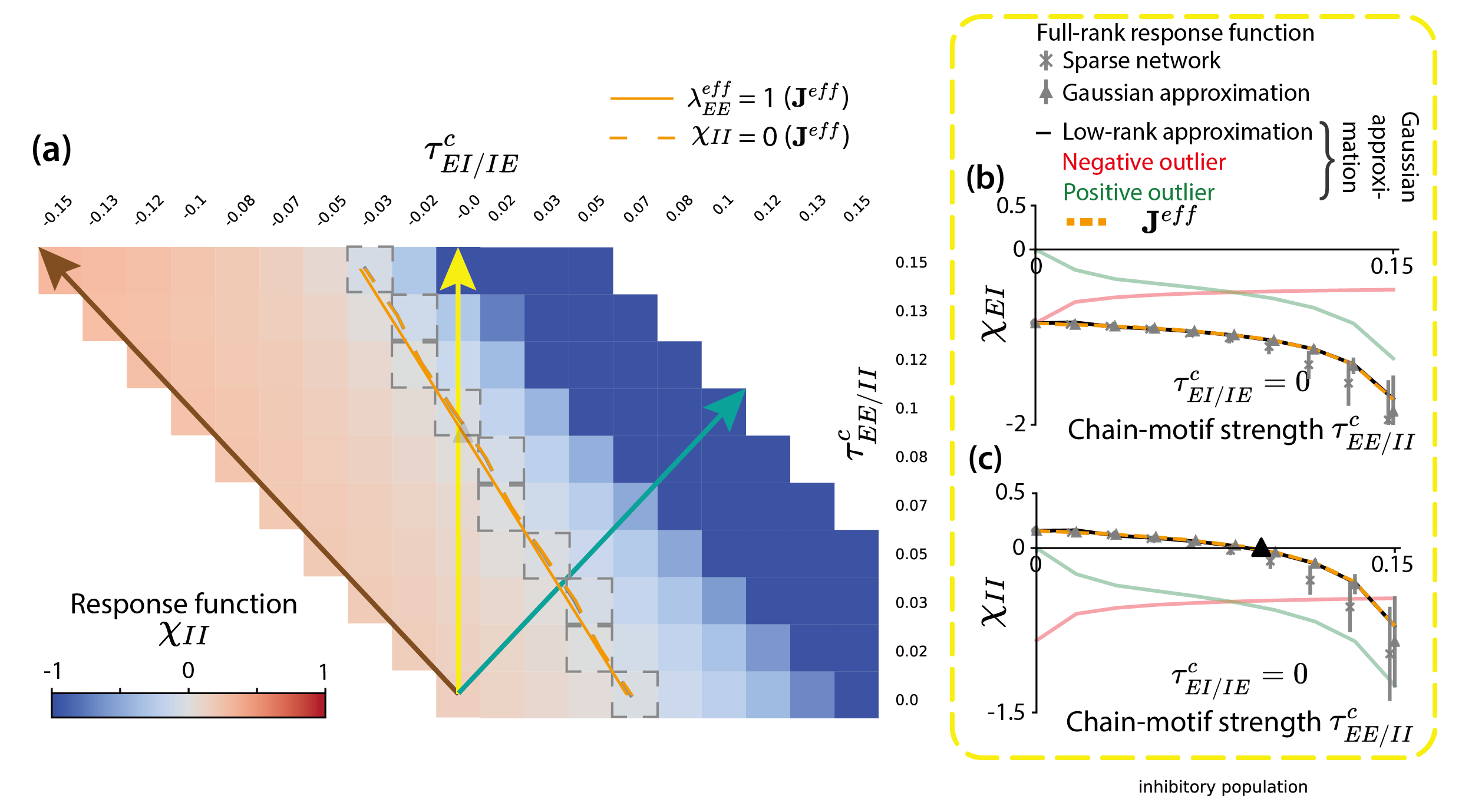}
\caption{\label{fig:betweenGaussian_Sparse}
Impact of the heterogeneity of chain motifs on the response to external inputs. (a) Response $\chi_{II}$ of inhibitory neurons as a function of $\tau^c_{EE/II}=\tau^c_{EE}=\tau^c_{II}$ and $\tau^c_{EI/IE}=\tau^c_{EI}=\tau^c_{IE}$, for the fully connected Gaussian approximation of sparse networks. The green line corresponds to homogeneous chain motifs ($\tau^c_{EE}=\tau^c_{II}=\tau^c_{EI}=\tau^c_{IE}$), for which the response is shown in \textcolor{teal}{\bf Fig~\ref{fig:GaussianParadoxical_nonISN}} . The 
brown line corresponds to homogeneous chain motifs occurence $\rho^C$, leading to $\tau^c_{EE}=\tau^c_{II}=-\tau^c_{EI}=-\tau^c_{IE}$, for which the response is shown in \textcolor{brown}{\bf Fig~\ref{fig:SparsenonISN_InhInhInp}}. 
The yellow line shows the prediction for networks with $\tau_{EI/IE}^c = 0$. 
Coordinates (squares) with a dashed border indicate the first value of $\tau_{EI/IE}^c$ that results in a paradoxical effect as $\tau_{EI/IE}^c$ gradually increases, while under a fixed $\tau_{EE/II}^c$. The orange dashed line represents the theoretical prediction for the boundary between the paradoxical and the non-paradoxical regime using $\mathbf{J}^{eff}$ and Eq.~\eqref{eq:response_full_rank}, the orange solid line represents the prediction using the stability of the E-E subnetwork $\mathbf{J}^{eff}_{EE}$ (Eq.~\eqref{eq:stabilityEEsubnet})
 (b) Response function $\chi_{EI}$ and (c)  response function  $\chi_{II}$ in sparse networks and corresponding fully connected Gaussian approximations with $\tau_{EI/IE}^c = 0$.
 Details of subplots (b) and (c) are identical to  Fig~\ref{fig:SparsenonISN_InhInhInp}.
 The other network parameters are identical to Fig~\ref{fig:SparsenonISN_InhInhInp}.
}
\end{figure*}

\subsection{\label{sub:HeterogeneousParadoxical}Effects of chain motif heterogeneity on paradoxical responses} 
To understand the origin of seemingly opposite effects of chain motifs on paradoxical responses in sparse and fully-connected networks, we note that in fully-connected networks we used homogeneous, positive $\tau^c$, while 
for sparse networks with a uniform chain motif occurrence $\rho^c$, the correlation coefficients are different across blocks of the connectivity matrix, with $\tau_{EE}^c=\tau_{II}^c$ positive and $\tau_{EI}^c=\tau_{IE}^c$ negative. We hypothesized that this difference may lead to the distinct dependences of the responses $\chi_{pq}$ to the strength of chain motifs. To investigate this possibility, we computed the responses $\chi_{II}$ in Gaussian networks where the chain motif strengths were not homogeneous across blocks of the connectivity matrix, but were instead determined by two parameters $\tau_{EE/II}^c$ and $\tau_{EI/EI}^c$ such that $\tau_{EE}^c=\tau_{II}^c=\tau_{EE/II}^c$ and $\tau_{EI}^c=\tau_{IE}^c=\tau_{EI/EI}^c$ (Fig~\ref{fig:betweenGaussian_Sparse}). This analysis reconciled the findings in Fig~\ref{fig:GaussianParadoxical_nonISN} and Fig~\ref{fig:SparsenonISN_InhInhInp}: increasing jointly $\tau_{EE/II}^c=\tau_{EI/IE}^c$ leads to a decreasing response $\chi_{II}$(green  line in Fig~\ref{fig:betweenGaussian_Sparse}), while setting $\tau_{EE/II}^c=-\tau_{EI/IE}^c$ instead leads to an increase of $\chi_{II}$ with the strength of chain motifs (
brown line in Fig~\ref{fig:betweenGaussian_Sparse}). 
We moreover predicted that setting $\tau_{EI/IE}^c=0$ and increasing $\tau_{EE/II}^c$ would lead to a decreasing $\chi_{II}$, and therefore a switch from a non-paradoxical to a paradoxical response (yellow  line and frame in Fig~\ref{fig:betweenGaussian_Sparse}). We directly verified this prediction in sparse networks by setting  $\rho^c_{EI/IE}=c^2$ and therefore zero correlation coefficients $\tau_{EI/IE}^c=0$, while maintaining $\tau_{EE/II}^c=\tau^c>0$ and gradually increasing this value (Fig~\ref{fig:betweenGaussian_Sparse}(b) and (c)). 

We can alternatively use the effective connectivity to examine paradoxical responses in networks with heterogeneous  variances and chain motifs. In this more general case, the explicit expression for $\mathbf{J}^{eff}$ is given by
\begin{equation}\label{eq:J_eff_heterogeneous}
\begin{aligned}
    J_{ij}^{eff} 
        =\left \{
        \begin{matrix}
         cJ+\alpha_E\sigma_E^2\tau_{EE}^c+\alpha_I\sigma_I\sigma_E\tau_{IE}^c,&j\in E\\
        -cgJ + \alpha_E\sigma_E\sigma_I\tau_{EI}^c+\alpha_I\sigma_I^2\tau_{II}^c,&j\in I
        \end{matrix}
        \right.
\end{aligned}
\end{equation}
We find that this effective connectivity, predicts accurately the response functions in the full, sparse network (orange dashed lines in Figs.~\ref{fig:dynamicsExternalInpAdjacency},~\ref{fig:SparsenonISN_InhInhInp} and \ref{fig:betweenGaussian_Sparse}(b)).

We next investigated to which extent the stability of the E-E subnetwork of $\mathbf{J}^{eff}$ predicts
 paradoxical responses of the inhibitory population in the full network. The connectivity of the E-E subnetwork is given by
\begin{equation}
\mathbf{J}_{EE}^{eff} = \left(cJ+\alpha_E\sigma_E^2\tau_{EE}^c+\alpha_I\sigma_I\sigma_E\tau_{IE}^c\right)\boldsymbol{e}\boldsymbol{e}^{\intercal},\,\,\,\boldsymbol{e}\in\mathbb{R}^{N_E}
\end{equation}
and has only one non-zero eigenvalue:
\begin{equation}\label{eq:stabilityEEsubnet}
\lambda_{EE}^{eff} = N_E(cJ+\alpha_E\sigma_E^2\tau_{EE}^c+\alpha_I\sigma_I\sigma_E\tau_{IE}^c).
\end{equation}
For the sparse network with a uniform occurrence $\rho^c$ we have $\tau^c = \tau_{EE/II}^c = -\tau_{EI/IE}^c$ (Figs.~\ref{fig:dynamicsExternalInpAdjacency} and \ref{fig:SparsenonISN_InhInhInp}), so that the eigenvalue is 
\begin{equation}
\lambda_{EE}^{eff} = N_EcJ+\alpha_E^2N^2J^2c(1-c)\tau^c(1-g\gamma).
\end{equation}
With an initial condition of $ N_E c J < 1 $ and an inhibition-dominant network $( 1 - g\gamma < 0 )$, the eigenvalue remains negative as  $\tau^c$ is increased, predicting correctly that the response of the inhibitory population remains non-paradoxical. 

However, when setting $\tau_{EI/IE}^c = 0$, the eigenvalue takes the form  
$\lambda_{EE}^{eff} = N_E c J + \alpha_E^2 N^2 J^2 c (1 - c) \tau_{EE/II}^c$,
and leads to a paradoxical inhibitory response as $\tau_{EE/II}^c$ is increased (Fig.~\ref{fig:betweenGaussian_Sparse}(b), black triangle).  
Furthermore, this stability criterion for the E-E subnetwork accurately predicts the transition boundary between paradoxical and non-paradoxical responses in the $\tau_{EE/II}^c -\tau_{EI/EI}^c$ plane (Fig.~\ref{fig:betweenGaussian_Sparse}(a)).

More generally, in the two-population networks we consider, chain motifs are specified by four independent parameters $\tau_{EE}^c, \tau_{II}^c, \tau_{EI}^c$ and $\tau_{IE}^c$. Figs.~\ref{fig:Supp_celltype_contributions}-\ref{fig:Supp_celltype_neg_contribution} illustrate the effects on the low-rank approximation of varying each of these parameters independently. In particular, negative values of chain motifs can induce pairs of complex conjugate eigenvalues.

\section{\label{secs:Discussion}Discussion}

Our findings demonstrate that chain motifs, a particular type of correlations between weights of pairs of synapses, exert a dominant impact on the dynamics in networks of recurrently connected neurons. Specifically, our mathematical analyses reveal that even a weak over-representation of chain motifs leads to the emergence of isolated eigenmodes distinct from networks with uncorrelated synaptic weights.
In inhibition-dominated networks, these additional eigenmodes are associated with a positive eigenvalue that reflects positive feedback induced by chain motifs. This positive feedback leads to a potential instability that requires revisiting the classical balance conditions.

The positive eigenmode generated by chain motifs strongly contributes to population-averaged responses to external inputs. In particular, we show that through this eigenmode, chain motifs can induce paradoxical responses in networks where such responses are not expected based on mean connectivity alone \cite{tsodyks1997paradoxical,miller2020generalized,sanzeni2020inhibition,garcia2017paradoxical,sadeh2020patterned}. Conversely, in inhibition-stabilized networks that exhibit paradoxical responses in the absence of correlations between synapses, chain motifs can shift the dynamics towards non-paradoxical responses. Mean connectivity therefore does not predict the nature of the responses when chain motifs are present.
Remarkably, we instead find that the response to external inputs is accurately predicted by a deterministic effective matrix $\mathbf{J}^{eff}$ (Eq.~\eqref{eq:Jeff_def}), in which the mean synaptic strengths are corrected by the strength of chain and reciprocal motifs. In particular, paradoxical responses correspond to networks where the effective excitatory sub-network defined by $\mathbf{J}^{eff}$ is unstable by itself \cite{miller2020generalized}.

Our results for the impact of chain motifs on responses to inputs were therefore obtained by comparing two approaches: (i) a low-rank approximation based on the outliers in the spectrum of the full, random connectivity matrix $\mathbf{J}$; (ii) an approximation of the response using an effective deterministic connectivity $\mathbf{J}^{eff}$. The two approaches provide complementary pictures and have different strengths and weaknesses. The low-rank approximation can in principle be applied to networks with arbitrary connectivity structure. It preserves the dominant eigenvalues of the full connectivity, and is therefore expected to capture transient dynamics. However, it relies on the knowledge of the statistics of eigenvectors, which  are in general not accessible analytically. The effective connectivity approach instead directly maps a large, random network with correlated synapses onto a classical deterministic network where each population is represented by one variable. A key difference with previous studies is that the connectivity weights in the effective deterministic network correspond to combinations of mean connectivity and motif strengths in the original random network.  It is important to note that the effective connectivity matrix does not accurately capture the outliers in the eigenspectrum of the full connectivity matrix (Appendix \ref{ap:EffectiveConnectivityMatrix}). While it predicts correctly the steady-state response, in general the effective connectivity is not expected to reproduce transient dynamics. Moreover, its form is specific to a connectivity structure based on pair-wise motifs.

An over-representation of chain motifs means that neurons with stronger inputs also have stronger outputs than expected by chance. This is precisely the case in the fully connected networks we investigated, where synaptic weights are continuous. In sparse networks, all synapses of a given type have the same strength, hence an over-representation of chain motifs means that neurons that receive more incoming synapses send out more outgoing synapses. This pattern is related, but not equivalent to network hubs. The concept of ‘hubs’ used in network science is generally based on the degree of individual nodes in a network \cite{van2013network}.
In many examples, including biological neuronal networks, the degree distribution is heavy-tailed, meaning that a small number of nodes exhibit disproportionately high connectivity, and such nodes are called ‘hubs’. An over-representation of chain motifs implies a heavy-tailed degree distribution and therefore the existence of hub neurons with both high in-degree and high out-degree.

Previous works have examined the effects of connectivity motifs on the continuous bulk of the spectrum \cite{kuczala2016eigenvalue,hu2022spectrum,dahmen2020strong,marti2018correlations,layer2024effect}. Here we instead focus on eigenmodes corresponding to eigenvalues that lie outside of the bulk. While the abundance of all four motif types along with the variance were found to significantly impact the spectral radius of the bulk  \cite{dahmen2020strong,hu2022spectrum}, our results show that connectivity motifs have very distinct effects on the outliers, with chain motifs most dominantly shaping their behavior and determining the steady-state responses to external inputs.
 
Each connectivity eigenvalue is associated to a corresponding eigenvector and shapes the dynamics in this direction in the N-dimensional space of neurons. Since eigenvalues are densely packed in the connectivity bulk spectrum, many of these directions a priori play an important role in shaping the collective dynamics of neural circuits. However, when the spectral radius of the bulk approaches the line of instability at $Re(\lambda)=1$, the most positive eigenvalues close to instability dominate the dynamics, yielding a dimensionality of neural activity that is much lower than the number of neurons in the circuit \cite{dahmen2020strong}. The impact of a mode in the bulk to the overall network activity thereby continuously degrades with distance to the instability line, as reflected by smooth covariance spectra \cite{hu2022spectrum}. Outlier eigenvalues instead each determine a one-dimensional activity component that is weighing much differently for the overall network behavior due to the separation to the other eigenvalues. Their associated dynamics can thus be captured by a low-rank approximation and identified by a larger gap in the covariance spectrum.

The present paper, with its emphasis on spectral properties of networks, complements allied studies on the role of connection motifs on network dynamics \cite{marti2018correlations,dahmen2020strong, hu2022spectrum, shao2023relating,zhao2011synchronization, hu2018feedback,hu2022spectrum,dahmen2020strong,trousdale2012impact,hu2013motif,hu2014local,ocker2017linking,recanatesi19dimension}.  Perhaps most related to the present paper is~\cite{hu2018feedback}, which expresses the response function for similar linearized rate networks in terms of a sequence of motif cumulants, which are the occurrences of  motif structures at levels over and above what one would expect from the occurrence of their constituent parts.  This leads to a general strong role for chains in particular over and above other motifs.  The present paper shares this emphasis on the importance of chain motifs, but arrives at this conclusion via different methods based on explicit computations of eigenvalues based on correlations among connections rather than the cumulant-based techniques of~\cite{hu2018feedback}.  Moreover, here we focus on distinct questions:  the impact of chain motifs on stability and paradoxical responses in the ubiquitous setting of balanced E-I networks, and the general interplay among synaptic motifs and larger-scale, low-rank connectivity patterns established by cell-type specific connectivity probabilities.

For the sake of analytical tractability, our study is based on a number of simplifying assumptions. We considered only linear, rate-based dynamics, while previous works outlined the effects of the non-linearity on paradoxical responses, in particular in supralinear-stabilized networks of excitatory and inhibitory neurons \cite{ahmadian2013analysis,rubin2015stabilized,litwin2016inhibitory,wu2023inhibition,holt2024stabilized}. Although a full non-linear analysis of N-dimensional networks with correlated synapses is in general untractable, networks with low-rank connectivity allow for an exact low-dimensional reduction even in the presence of non-linearities \cite{mastrogiuseppe2018linking, beiran2021shaping, dubreuil2022role}. Applying the low-rank approximation of correlated connectivity to supralinear stabilized networks is therefore an interesting direction for future research. Beyond rate-based models, numerical simulations of integrate-and-fire networks with low-rank connectivity structure suggest that the analytical insights from rate networks extend to spiking networks \cite{cimevsa2023geometry}.

In this work, we focused on networks of only two populations, with the additional constraint that the mean connectivity $\mathbf{J^0}$ is unit rank. As a consequence, the mean connectivity induces a single eigenmode, which in inhibition-dominated networks corresponds to a negative outlying eigenvalue. Our key result is that in this situation chain motifs lead to an additional eigenmode with a positive outlying eigenvalue. Relaxing the unit-rank constraint, and increasing the number of populations leads to additional non-trivial eigenmodes of the mean connectivity $\mathbf{J^0}$, and additional outlying eigenvalues induced by chain motifs. Systematically mapping the resulting eigenspectrum configurations is left for future work. It would be particularly interesting to examine the effects of multiple inhibitory sub-types and their interactions with synaptic motifs.

How large is a given value of $\tau^c$? To get an intuitive understanding, it is useful to focus on sparse networks. In that case, the correlation coefficient $\tau^c$ is directly related to the occurrence probability $\rho^c$ (Eq.~\eqref{eq:abundance2Covariance}, which corresponds to the fraction of chain motifs observed across all possible chain motifs.
A value of  $\tau^c = 0.1$ together with the connectivity probability $c = 0.2$, corresponds to an occurrence probability of $\rho^c = 0.056$. This is only $0.016$ above the chance level given by $c^2 = 0.04$. This implies that in an experiment where we examine $1000$ neuron triplets that could potentially form a chain motif, only about $56$ triplets will have the pair of synapses connected. Among these, $40$ triplets are expected purely due to chance, so only $16$ triplets contribute to an over-representation beyond chance. This level of over-representation therefore corresponds to a relatively small number of chain motifs, and yet has a substantial impact on the network dynamics. 

Experimental studies of optogenetic perturbations have revealed the presence of paradoxical inhibitory responses throughout the cortex \citep{sanzeni2020inhibition, palmigiano2020common, sanzeni2023mechanisms}. One proposed explanation is that these paradoxical responses stem from strong average recurrent excitation. Our findings show that, alternatively, paradoxical responses could be due to a weak over-representation of chain motifs. In recent physiological datasets, the measured strength of chain motifs is in the range of $\tau^c=0.05-0.1$ \cite{dahmen2020strong}, and our results show that these values are compatible with paradoxical responses even if average recurrent excitation is weak. Additional investigations will however be needed to distinguish experimentally between these two possible mechanisms.

\section*{Data availability}
The code for all simulations and figure generation will be available on GitHub upon publication of the paper.

\begin{acknowledgments}
YS and SO were supported by the Eranet-Neuron project IMBALANCE and the program “Ecoles Universitaires de Recherche” launched by the French Government and implemented by the ANR, with the reference ANR-17-EURE-0017. YS was supported by the National Natural Science Foundation of China (grant no. 32400936), the Fundamental Research Funds for the Central Universities (grant no. 2233100030), the NSF AccelNet IN-BIC program (grant No. OISE-2019976). DD was funded by the Deutsche Forschungsgemeinschaft (DFG, German Research Foundation) as part of the SPP 2205 – 533396241. We additionally acknowledge partial support of NIH grant R01 1RF1DA055669. We thank K.D. Miller for feedback on the manuscript and the suggestion to look for an equivalent effective matrix.
\end{acknowledgments}

\cleardoublepage
\appendix
\counterwithin{figure}{section}
\begin{table*}
    \caption{\label{tab:parameters}List of simulation parameters.}
    \begin{ruledtabular}
    \begin{tabular}{ccccccc}
    \textrm{Parameters}&
    \multicolumn{5}{c}{\textrm{Simulations}}\\
    \colrule
&Figs~\ref{fig:F1_ConnStatsOutliers},\ref{fig:comparable_ReciprocalMotifs}(a-c)&  Figs~\ref{fig:F1_ConnStatsVectors},\ref{fig:dynamicsExternalInp}, & Fig~\ref{fig:comparable_ReciprocalMotifs}(d)&Figs~\ref{fig:EigenvaluesAdjacencyMatrix},&Figs~\ref{fig:AdjacencymatrixVectors},\ref{fig:dynamicsExternalInpAdjacency},&Figs~\ref{fig:SparsenonISN_InhInhInp},\ref{fig:betweenGaussian_Sparse}\\
     &&\ref{fig:ReciprocalDyns}(a)& &\ref{fig:comparable_ReciprocalMotifs}(e)&\ref{fig:ReciprocalDyns}(b),\ref{fig:SparseISN_2_NonISN}&\ref{fig:Supp_AdjacencymatrixVectors}\\
     \colrule
     $N$ & 1000 & 1000 & 1000 & 1500&1500&1500  \\
      $J^0$ & 8.125$\times$ 1E-4& 8.125$\times$ 1E-4& 8.125$\times$ 1E-4& \-- & \--& \--  \\
      $\sigma$& 0.2& 0.1& 0.5&\--&\--&\-- \\
      $g$&10.15 &10.15 &10.15 &6.8&6.0&6.0\\
      $\gamma$ & 1/4& 1/4 & 1/4  &1/4&1/4&1/4\\
      $c$ &\--&\--&\--& 0.2& 0.2& 0.2\\
      $J$&\--&\--&\-- &0.0129&0.0129&0.00325\\
     
\end{tabular}
\end{ruledtabular}
\end{table*}

\section{\label{ap:GeneralMotifs} Other second-order motifs}
In the main text, we focus on the chain and reciprocal motifs, however two other types of second-order motifs can be distinguished: divergent and convergent motifs (Fig.~\ref{fig:Schematic}(c)).
Divergent motifs correspond to the correlation coefficient $\tau^{div}_{ikj}$ between synapses $J_{ij}$ and $J_{kj}$ which share an identical pre-synaptic neuron $j$ but have different post-synaptic neurons $i\neq k$:
\begin{equation}\label{eq:divergent_ap}
\begin{aligned}
\tau^{div}_{ikj} = \frac{[J_{ij}J_{kj}]-[J_{ij}][J_{kj}]}{\sqrt{[(J_{ij}-[J_{ij}])^2][(J_{kj}-[J_{kj}])^2]}}=\frac{[z_{ij}z_{kj}]}{\sqrt{[z_{ij}^2][z_{kj}^2]}};
\end{aligned}
\end{equation}
Convergent motifs correspond to the correlation coefficient $\tau^{con}_{ikj}$ betwen synapses $J_{ik}$ and $J_{ij}$ which share the identical post-synaptic neuron $i$ but have different pre-synaptic neurons $j\neq k$:
\begin{equation}\label{eq:convergent_ap}
\begin{aligned}
\tau^{con}_{ikj} = \frac{[J_{ik}J_{ij}]-[J_{ik}][J_{ij}]}{\sqrt{[(J_{ik}-[J_{ik}])^2][(J_{ij}-[J_{ij}])^2]}}=\frac{[z_{ik}z_{ij}]}{\sqrt{[z_{ik}^2][z_{ij}^2]}}.
\end{aligned}
\end{equation}
Only chain and reciprocal motifs contribute to the dynamics for large $N$ because they correspond to  elements of the matrix $[\mathbf{Z}^2]$ (Eq.~\eqref{eq:motifQuadraticZ}) that determines the outlying eigenvalues.
In contrast, divergent and convergent motifs do not affect the matrix $[\mathbf{Z}^2]$, but they contribute to higher powers $[\mathbf{Z}^{2k}]$ in a sub-dominant fashion.
To illustrate this higher-order impact, we consider a single population network with homogeneous motif strengths $\tau^c,\tau^r,\tau^{div},\tau^{con}$ and a homogeneous variance of $\mathbf{Z}$, denoted by $\sigma_z^2$, and calculate $[\mathbf{Z}^4]$ as an example. Computing the product of two entries $\left(\mathbf{Z}^2\right)_{ij},~\left(\mathbf{Z}^2\right)_{jk},~i\neq j\neq k$, its average is expressed as
\begin{widetext}
\begin{equation}
\begin{aligned}
\left[\left(\mathbf{Z}^2\right)_{ij}\left(\mathbf{Z}^2\right)_{jk}\right] &= \sum_{s,l}^{N}\left[z_{is}z_{sj}z_{jl}z_{lk}\right]\\
&=\sum_{s,l}^N\big(\left[z_{is}z_{sj}\right]\left[z_{jl}z_{lk}\right]+\left[z_{is}z_{jl}\right]\left[z_{sj}z_{lk}\right]
+\left[z_{is}z_{lk}\right]\left[z_{jl}z_{sj}\right]\big)\\
&=\left[\mathbf{Z}^2\right]_{ij}\left[\mathbf{Z}^2\right]_{jk} 
+\sum_{s}^{N}\left[z_{is}z_{js}\right]\left[z_{sj}z_{sk}\right]
+ \sum_{s}^{N}\left[z_{is}z_{sk}\right]\left[z_{js}z_{sj}\right]\\
&=\left[\mathbf{Z}^2\right]_{ij}\left[\mathbf{Z}^2\right]_{jk} 
+ N\tau^{con}\sigma_z^2\tau^{div}\sigma_z^2
+ N\tau^{c}\sigma_z^2\tau^{r}\sigma_z^2.
\end{aligned}
\end{equation}
\end{widetext}
We used Wick's theorem from the first line to the second line. From the second line to the third line, we used $i\neq j\neq k$. We moreover assumed $\left[z_{is}z_{jl}\right]=0$ unless two of the pre-synaptic neuron indices are identical, so 
that the second and third terms are non-zero only when $s=l$.
Since (Eq.~\eqref{eq:motifQuadraticZ})
\begin{equation}
\left[\mathbf{Z}^2\right]_{ij} = N\tau^{c}\sigma_z^2,\,\,\,\,i\neq j
\end{equation}
we have, 
\begin{equation}
\begin{aligned}
\frac{\left[\left(\mathbf{Z}^2\right)_{ij}\left(\mathbf{Z}^2\right)_{jk}\right]}{\left[\mathbf{Z}^2\right]_{ij}\left[\mathbf{Z}^2\right]_{jk}}&=1+\frac{N\tau^{div}\tau^{con}\sigma_z^4+N\tau^{c}\tau^{r}\sigma_z^4}{(N\tau^{c})^2\sigma_z^4}\\
&=1+ \mathcal{O}(N^{-1}).
\end{aligned}
\end{equation}
The correlation between fluctuations in $\left(\mathbf{Z}^2\right)_{ij}$ and $\left(\mathbf{Z}^2\right)_{jk}$ diminishes with a scaling factor of $1/N$. This correlation tends to vanish in the limit of a large network as $N$ approaches infinity. 

\section{\label{ap:EigenvalueCalculations} Eigenvalue calculations}
\subsection{\label{ap:nonSelfAveraging} General approach}
Using the determinant lemma (Eq.~\eqref{eq:MeanPlusRandom}-\eqref{eq:determinantLemma}) we show that the eigenvalues of matrix $\mathbf{J}$ after perturbations are the solutions of the characteristic polynomial of the matrix $\mathbf{Q}$ (Eq.~\eqref{eq:seriesRepresentation}). 
As $[\mathbf{Z}^k]=0$ for odd $k$, the realization average of $\mathbf{Q}$ is given by

\begin{equation}\label{eq:ensembleAvgZ}
\left[\mathbf{Q}\right] = \sum_{l=0}^{\infty}\mathbf{N}_0^{\intercal}\left [ \mathbf{Z}^{2l}\right ]\mathbf{M}_0/(N\lambda^{2l})
\end{equation}

In general, $[\mathbf{Z}^{2l}]\neq [\mathbf{Z}^2]^l$, however,  the correlation between entries $(\mathbf{Z}^2)_{ij}$ and $(\mathbf{Z}^2)_{jk}$ vanishes in the large network limit $N\to \infty$ (Appendix ~\ref{ap:GeneralMotifs}).

Consequently, we rewrite the expression 
\begin{equation}
\left[\mathbf{Q}\right] = \sum_{l=0}^{\infty}\mathbf{N}_0^{\intercal}\left [ \mathbf{Z}^2\right ]^l\mathbf{M}_0/(N\lambda^{2l}).
\end{equation}
Using the geometric sequence summation calculation, we get Eq.~\eqref{eq:OutlierSecondorder} in the main text
\begin{equation}
    \begin{aligned}
    \left[\mathbf{Q}\right] &=\frac{1}{N}\mathbf{N}_0^{\intercal}\left(\mathbf{1}-\frac{\left[\mathbf{Z}^2\right]}{\lambda^2}\right)^{-1}\mathbf{M}_0.
    \end{aligned}
\end{equation}
The matrix $[\mathbf{Z}^2]$ can be decomposed as $\mathbf{D}+\mathbf{O}$, where $\mathbf{D}\in\mathbb{R}^{N\times N}$ is a diagonal matrix, and $\mathbf{O}\in\mathbb{R}^{N\times N}$ is a matrix of identical elements with block structure.
The entries in matrix $\mathbf{D}$ are determined by both the reciprocal and chain strengths, and the entries in matrix $\mathbf{O}$ are determined by the chain motif strength.
Furthermore, the matrix $\mathbf{O}$ can be decomposed and expressed as 
\begin{equation}
    \mathbf{O}=\mathbf{U}_o\mathbf{V}_o^{\intercal};\,\,\,\,\,\,\,\
    \mathbf{U}_o,~\mathbf{V}_o\in\mathbb{R}^{N\times R_o}
\end{equation} 
where the  rank of $\mathbf{O}$ is $R_o\ll P$. 
Then, we use the Woodbury matrix identity: given a square invertible $N\times N$ matrix $\mathbf{1}-\mathbf{D}/\lambda^2$, two $N\times R_o$ matrices $\mathbf{U}_o$ and $\mathbf{V}_o$, $\mathbf{A}$ and $\mathbf{B}$ the invertible $N\times N$ matrices 
\begin{eqnarray}
    \mathbf{A}&=&\left(\mathbf{1}-\frac{\mathbf{D}}{\lambda^2}\right)={\rm diag}\left(\left\{\frac{\lambda^2-D_{ii}}{\lambda^2}\right\}\right)\\
    \mathbf{B}&=&\left(\mathbf{1}-\frac{\left[\mathbf{Z}^2\right]}{\lambda^2}\right)=\left(\left(\mathbf{1}-\frac{\mathbf{D}}{\lambda^2}\right)-\frac{\mathbf{U}_o\mathbf{V}_o^{\intercal}}{\lambda^2}\right)
\end{eqnarray}
we have 
\begin{equation}\label{eq:inverseB}
\begin{aligned}
\mathbf{B}^{-1} &= \mathbf{A}^{-1}+\frac{1}{\lambda^2}\mathbf{A}^{-1}
\mathbf{U}_o\left(\mathbf{1}_{R_o}-\frac{1}{\lambda^2}\mathbf{V}_o^{\intercal}\mathbf{A}^{-1}\mathbf{U}_o\right)^{-1}\mathbf{V}_o^{\intercal}\mathbf{A}^{-1}.
\end{aligned}
\end{equation}
Now, rather than computing the inverse of an $N\times N$ matrix $(\mathbf{1}-\mathbf{Z}/\lambda)$ (Eq.~\eqref{eq:determinantLemma}), we calculate the inverse of a diagonal matrix $\mathbf{A}$ and an $R_o\times R_o$ matrix $(\mathbf{1}_{R_o}-\mathbf{V}_o^{\intercal}\mathbf{A}^{-1}\mathbf{U}_o/\lambda^2)$. 

We substitute Eq.~\eqref{eq:inverseB} into Eq.~\eqref{eq:OutlierSecondorder}, and then search for the solutions of the characteristic polynomial $f_Q(\lambda)=0$ (Eq.~\eqref{eq:OutlierInfinite}) to obtain the eigenvalue outliers $\lambda_r,~r=1\dots R$.

\subsection{\label{ap:homogeneousGauss}Gaussian networks with uniform variance and correlations}

In this section, we calculate the eigenvalue outliers in Gaussian networks with homogeneous chain motif strength (Sec.~\ref{subsecs:EigenvalueGauss}). We consider a homogeneous variance parameter $\sigma^2$ (Eq.~\eqref{eq:scaledvariance_homo}) and correlation coefficient of chain motifs $\tau^c$, and the elements of $[\mathbf{Z}^2]$ are given by 
\begin{equation}\label{eq:motifQuadraticZ_chn_ap}
\left[\sum_{k=1}^Nz_{ik}z_{kj}\right] =\left \{
\begin{matrix}
\sum_{q=1}^P \alpha_q\sigma^2\tau^c,&i\neq j,\\
 0,&i=j.\
\end{matrix}
\right.
\end{equation}
In Eq.~\eqref{eq:ZsquareDecompose_Maintext} we then have
\begin{eqnarray}\label{eq:homogeneousDO}
\mathbf{D} &=& -\tau^c\sigma^2\mathbf{1}\\
\quad\mathbf{O} &=& {\tau^c\sigma^2}\boldsymbol{e}\boldsymbol{e}^{\intercal}
\end{eqnarray}
where $R_o=1$ with $\mathbf{U}_o=\boldsymbol{e}$, $\mathbf{V}_o=\tau^c\sigma^2\boldsymbol{e}$, and $\boldsymbol{e}$ is an all-one vector. Substituting Eq.~\eqref{eq:homogeneousDO} into Eq.~\eqref{eq:inverseB}, we obtain
\begin{equation}\label{eq:homogeneousInverseB}
\mathbf{B}^{-1} = \frac{\lambda^2}{\lambda^2+\tau^c\sigma^2}\mathbf{1}+\frac{\lambda^2\tau^c\sigma^2}{\left(\lambda^2-(N-1)\tau^c\sigma^2\right)\left(\lambda^2+\tau^c\sigma^2\right)}\boldsymbol{e}\boldsymbol{e}^{\intercal}.
\end{equation}
Combining the rank-one mean connectivity structures as described in Eq.~\eqref{eq:meanstructures_Gaussian} and inserting this $\mathbf{B}^{-1}$ (from Eq.~\eqref{eq:homogeneousInverseB}) into Eq.~\eqref{eq:OutlierSecondorder} and subsequently into Eq.~\eqref{eq:OutlierInfinite}
\begin{equation}
\begin{aligned}
\lambda &= \frac{1}{N}\boldsymbol{n}_0^{\intercal}\mathbf{B}^{-1}\boldsymbol{m}_0\\
\boldsymbol{m}_0 &= [1\dots]^{\intercal} = \boldsymbol{e}\\
\boldsymbol{n}_0 &= [NJ^0\dots,-NgJ^0\dots]^{\intercal}
\end{aligned}
\end{equation}
we obtain the polynomial equation for $\lambda$:
\begin{equation}\label{eq:polynomial_EigenvalueGaussian_supp}
\lambda = \frac{\lambda^2\lambda_0}{\lambda^2+\tau^c\sigma^2}+\frac{N\lambda^2\tau^c\sigma^2\lambda_0}{\left(\lambda^2+\tau^c\sigma^2\right)\left(\lambda^2-(N-1)\tau^c\sigma^2\right)}.
\end{equation}
It's important to note that we focus on the case where $\lambda$  is an outlier, indicating $\lambda^2>\sigma^2$. Additionally, the correlation coefficient $\tau^c$ falls within the range $[-1,1]$, ensuring that $\lambda$ satisfies $\lambda\neq 0$ and $\lambda^2+\tau^c\sigma^2>0\neq 0$. 
These conditions lead us to the polynomial equation expressed in Eq.~\eqref{eq:polynomial_EigenvalueGaussian} in the main text 
\begin{equation}
\lambda=\frac{\lambda_0\mp\sqrt{\lambda_0^2+4(N-1)\tau^c\sigma^2}}{2}.
\end{equation}
Analogously, in Gaussian networks with homogeneous reciprocal motif strength, the elements of $[\mathbf{Z}^2]$ are given by 
\begin{equation}\label{eq:motifQuadraticZ_rec_ap}
\left[\sum_{k=1}^Nz_{ik}z_{kj}\right] =\left \{
\begin{matrix}
 0,&i\neq j,\\
\sum_{q=1}^P \alpha_q\sigma^2\tau^r,&i=j.\
\end{matrix}
\right.
\end{equation}
There is only the diagonal matrix $\mathbf{D}$ which has identical non-zero diagonal entries $\tau^r\sigma^2$, therefore 
\begin{equation}
\mathbf{B}^{-1}=\mathbf{A}^{-1}=\frac{\lambda^2}{\lambda^2-\tau^r\sigma^2}\mathbf{1}.
\end{equation}
Thus the polynomial equation for determining the eigenvalues of $\mathbf{J}$ with solely reciprocal motifs is 
\begin{equation}
\lambda = \frac{\lambda_0\lambda}{\lambda^2-\tau^r\sigma^2}.
\end{equation}
Given that $\lambda\neq 0$, we have
\begin{equation}
\lambda^2-\lambda\lambda_0-\tau^r\sigma^2=0,
\end{equation}
which leads to the Eq.~\eqref{eq:eigvperturbationStrength_rec_Gauss} in the main text and gives
\begin{equation}
\lambda = \frac{\lambda_0\mp\sqrt{\lambda_0^2+4\tau^r\sigma^2}}{2}.
\end{equation}

\subsection{\label{ap:heterogeneousGauss}Gaussian networks with heterogeneous variance and correlations}
Next, we derive the eigenvalues of a connectivity matrix with heterogeneous variance and correlations.
Rather than considering the general case where the variance of synaptic weights is determined by the populations of both pre- and postsynaptic neurons, we assume here that it depends solely on the population of the presynaptic neuron. Consequently, we define the variance as  $\sigma_{pq}^2=\sigma_{q}^2/N$.
For the correlation coefficients for chain motifs, we assume that they  depend on the populations of the presynaptic neurons of the two involved synapses, thereby introducing heterogeneity in $\tau_{pq}^c$.
Following these assumptions, the elements of  $[\mathbf{Z}^2]$ are  
\begin{equation}\label{eq:motifQuadraticZ_hetero_chn_ap}
\left[\sum_{k=1}^Nz_{ik}z_{kj}\right] =\left \{
\begin{matrix}
\sum_{q=1}^P \alpha_q\sigma_{q}\sigma_{s}\tau_{qs}^c,&i\neq j,\\
 0,&i=j,\
\end{matrix}
\right.
\end{equation}
where the neuron with index $i$ belongs to population $p$ while the neuron with index $j$ belongs to population $s$ with $p,q,s\in\{E,I\}$.  The matrix $[\mathbf{Z}^2]$ therefore has a columnar structure where all elements corresponding to $j\in E$ are identical, as are those for $j\in I$. 
Consequently, we express
\begin{equation}\label{eq:ZEI_square_hetero_ap}
\begin{aligned}
\left[\mathbf{Z}^2\right]_{ij} =& \alpha_E\sigma_E^2\tau_{EE}^c+\alpha_I\sigma_I\sigma_E\tau_{IE}^c =Z_E^2, j\in E\\
\left[\mathbf{Z}^2\right]_{ij} =& \alpha_E\sigma_E\sigma_I\tau_{EI}^c+\alpha_I\sigma_I^2\tau_{II}^c =Z_I^2, j\in I
\end{aligned}
\end{equation}
Analogously to the homogeneous case, the matrix $[\mathbf{Z}^2]$ can be decomposed into a diagonal matrix $\mathbf{D}$ and a block-like matrix $\mathbf{O}$. Since the impact of $\mathbf{D}$ on the eigenvalues is negligible compared to that of $\mathbf{O}$, for ease of mathematical analysis, we ignore $\mathbf{D}$ and focus only on the block-like matrix $\mathbf{O}$.
Moreover, as per Eq.~\eqref{eq:ZEI_square_hetero_ap}, the matrix $\mathbf{O}$ has a unit rank structure, expressed as 
\begin{equation}\label{eq:heterogeneousO}
\begin{aligned}
\mathbf{O} &= \mathbf{U}_o\mathbf{V}_o^{\intercal}\\
\mathbf{U}_o &= \boldsymbol{e}\\
\mathbf{V}_o &= \left[Z_E^2\dots,Z_I^2\dots\right]^{\intercal}.
\end{aligned}
\end{equation}
By substituting Eq.~\eqref{eq:heterogeneousO}, we have
\begin{equation}
\begin{aligned}
\mathbf{A} &= \mathbf{1}\\
\mathbf{B} &= (\mathbf{1}-\frac{1}{\lambda^2}\mathbf{U}_o\mathbf{V}_o^{\intercal})
\end{aligned}
\end{equation}
therefore 
\begin{equation}\label{eq:heterogeneousInverseB}
\mathbf{B}^{-1} = \mathbf{1}+\frac{1}{\lambda^2-\left(N_EZ_E^2+N_IZ_I^2\right)}\mathbf{U}_o\mathbf{V}_o^{\intercal}.
\end{equation}
By integrating the rank-one mean structure from Eq.\eqref{eq:meanstructures_Gaussian} with the inverse matrix described in Eq.\eqref{eq:heterogeneousInverseB} into Eq.~\eqref{eq:OutlierSecondorder}, we obtain the polynomial equation
\begin{equation}
\lambda = \lambda_0+\frac{\lambda_0\left(N_EZ_E^2+N_IZ_I^2\right)}{\lambda^2-\left(N_EZ_E^2+N_IZ_I^2\right)}.
\end{equation}
The non-zero solutions can be expressed as
\begin{equation}\label{eq:eigenvals_chn_ap}
\lambda= \frac{\lambda_0\mp\sqrt{\lambda_0^2+4\left(N_EZ_E^2+N_IZ_I^2\right)}}{2}
\end{equation}

Similarly, in Gaussian networks characterized by heterogeneous reciprocal motif strengths, the elements of $[\mathbf{Z}^2]$ are 
\begin{equation}\label{eq:motifQuadraticZ_hetero_rec_ap}
\left[\sum_{k=1}^Nz_{ik}z_{kj}\right] =\left \{
\begin{matrix}
 0,&i\neq j,\\
\sum_{q=1}^P \alpha_q\sigma_q\sigma_p\tau_{pq}^r,&i=j.\
\end{matrix}
\right.
\end{equation}
where the neuron with index $i$ belongs to population $p$.
In the simplified two-population scenario, the matrix $[\mathbf{Z}^2]$ only has non-zero diagonal elements. All diagonal elements corresponding to $i\in E$ are identical, as are those for $i\in I$. 
We denote these two values as 
\begin{equation}\label{eq:ZEI_square_hetero_rec_ap}
\begin{aligned}
\left[\mathbf{Z}^2\right]_{ii} =& \alpha_E\sigma_E^2\tau_{EE}^r+\alpha_I\sigma_I\sigma_E\tau_{IE}^r =Z_E^2, i\in E\\
\left[\mathbf{Z}^2\right]_{ii} =& \alpha_E\sigma_E\sigma_I\tau_{EI}^r+\alpha_I\sigma_I^2\tau_{II}^r =Z_I^2, i\in I.
\end{aligned}
\end{equation}
The inverse matrix $\mathbf{A}^{-1}={\rm diag}(\{\lambda^2/(\lambda^2-D_{ii})\})$ is therefore a diagonal matrix 
 \[ \begin{bmatrix}
   \frac{\lambda^2}{\lambda^2-Z_E^2} &  &  & \\ 
   & \ddots &  & \\ 
   &  &  \ddots & \\ 
   &  &   & \frac{\lambda^2}{\lambda^2-Z_I^2},
 \end{bmatrix} \]
 which, combined with the mean unit rank structure, leads to the polynomial equation for the eigenvalues as 
 \begin{equation}\label{eq:polynomial_rec_heter_ap}
 \lambda = \frac{N_EJ^0\lambda^2}{\lambda^2-Z_E^2}-\frac{N_IgJ^0\lambda^2}{\lambda^2-Z_I^2}.
 \end{equation}
By replacing $Z_E^2,~Z_I^2$ with Eq.~\eqref{eq:ZEI_square_hetero_rec_ap} and solving this polynomial equation, we can theoretically obtain the eigenvalues of $\mathbf{J}$ with heterogeneous reciprocal correlations \cite{shao2023relating}.

\section{\label{ap:networkconstruction}Network construction}

\subsection{\label{ap:networkconstructionGaussian} Fully-connected Gaussian networks}

To generate fully-connected networks with correlated Gaussian weights, we follow the standard procedure for generating correlated Gaussian variables by forming linear combinations of uncorrelated variables \cite{dahmen2020strong}.

For homogeneous networks we use
\begin{widetext}
\begin{equation}\label{eq:creationMotifConnectivity}
\begin{aligned}
    J_{ij}&=J^0_{pq}+\frac{\sigma}{\sqrt{N}}\big(\underbrace{\overbrace{sgn(\tau^c)\sqrt{\vert\tau^{c}\vert}\eta_i}^{\rm convergent}+\overbrace{\sqrt{\vert\tau^{c}\vert}\eta_j}^{\rm divergent}-sgn(\tau^c)\sqrt{\vert\tau^c\vert}\mu_{ij}+\sqrt{\vert\tau^c\vert}\mu_{ji}}_{\rm{chain}}\\
    &\underbrace{+sgn(\tau^r)\sqrt{\vert\tau^r/2\vert}\nu_{ij}+\sqrt{\vert\tau^r/2\vert}\nu_{ji}}_{\rm{reciprocal}}+\underbrace{\sqrt{1-4\vert\tau^c\vert-\vert\tau^r\vert}y_{ij}}_{\rm{independent}}\big)
\end{aligned}
\end{equation}
\end{widetext}
where the post- and presynaptic neurons $i$ and $j$ belong to population $p$ and $q$, $\tau^c$ denotes the homogeneous chain type correlations between connections in a chain circuit crossing neuron $i$ or $j$, $\tau^r$ represents the homogeneous reciprocal type correlation between connections $J_{ij}$ and $J_{ji}$, and $\eta_i,\mu_{ij},\nu_{ij}$, and $y_{ij}$ are i.i.d. normal random variables.

Note that, following  \cite{dahmen2020strong,hu2022spectrum}, we generate chain motifs by simultaneously creating divergent and convergent motifs (Appendix~\ref{ap:GeneralMotifs}) through the shared random variable $\eta$, so that  $\vert\tau^{div}\vert=\vert\tau^{con}\vert=\vert\tau^c\vert$. This is not the most general approach, as we could further add two independent random variables, $\alpha_i$ and $\beta_i$, for synaptic couplings that start or end at neuron $i$, respectively. These  random variables would independently contribute to $\tau^{div}$ and $\tau^{con}$, but not to $\tau^c$ or $\tau^r$. However, since only chain and reciprocal motifs  influence $[\mathbf{Z}^2]$ and the eigenvalues of the dynamics, we focus here on a simplified scenario where the divergent and convergent correlations fully characterize the chain motifs.

The terms $-sgn(\tau^c)\sqrt{\vert\tau^c\vert}\mu_{ij}+\sqrt{\vert\tau^c\vert}\mu_{ji}$ eliminate the diagonal reciprocity introduced by convergent and divergent motifs, 
the coefficient of the independent variable $y_{ij}$ imposes the constraint that $1-4\vert\tau^c\vert-\vert\tau^r\vert>0$.

For networks with population-dependent variances and correlations, we use
\begin{widetext}
\begin{equation}\label{eq:connectivityDescribtion}
\begin{aligned}
J_{ij} =& J^0_{pq}+{\sigma_{pq}}(sgn(\tau^c_{pq})\sqrt{\vert\tau^c_{pq}\vert}\eta_i^{(q)}+\sum_{x=1}^{P}\sqrt{\vert\tau^c_{qx}\vert}\eta_j^{(x)}-sgn(\tau^{c}_{pq})\sqrt{\vert\tau^{c}_{pq}\vert}\mu_{ij}^{(q)}+\sqrt{\vert\tau^c_{qp}\vert}\mu_{ji}^{(p)}\\
&+sgn(\tau^{r}_{pq})\sqrt{\vert\tau^{r}_{pq}/2\vert}\nu_{ij}+\sqrt{\vert\tau^r_{pq}/2\vert}\nu_{ji}+c_{rm,pq}y_{ij})\\
c_{rm,pq} =& \sqrt{1-2\vert\tau^c_{pq}\vert-\vert\tau^c_{qp}\vert-\sum_{x=1}^{P}\vert\tau^c_{qx}\vert-\vert\tau^r_{pq}\vert}
\end{aligned}
\end{equation}
\end{widetext}
where the pre-synaptic neuron with index $j$ belongs to population $q$, and the post-synaptic neuron with index $i$ belongs to population $p$. Here, $\tau^c_{pq}$ and $\tau^r_{pq}$ denote the population-dependent chain and reciprocal  correlation coefficients, while
$c_{rm,pq}$ is a  coefficient setting the magnitude of the independent part of the synaptic variance. The entries in vectors $\boldsymbol{\eta}^{(p)}\in \mathbb{R}^{N},~p=1\dots P$, matrices $\mathbf{U}^{(p)},~p=1\dots P$, $\mathbf{V}$ and $\mathbf{Y}\in \mathbb{R}^{N\times N}$ are i.i.d. random variables following the normal distribution $\mathcal{N}(0,1)$.

\subsection{\label{ap:Sparsenetwork}Sparse networks}
We use the SONET algorithm \cite{zhao2011synchronization} to generate sparse networks with predefined local motifs. In brief,  the occurrence probabilities of different motifs (Eq.~\eqref{eq:secondOrderSparse})  are determined by the in-degree, out-degree, and full, joint-degree  distributions of the units within the sparse networks. For each neuron in the network, SONET samples pairs of in- and out-degree values from the full-degree distribution $p_d(x,y)$, and then assigns a non-zero connection from neuron $j$ to $i$ with a probability proportional to $x_iy_j$, where the proportionality constant sets the sparsity in the network.

More specifically, the in-degree distribution $p_{in}(x)$ is associated with the convergent motifs probability $\rho^{con}$, the out-degree distribution $p_{out}(y)$ is associated with the divergent motifs probability $\rho^{div}$, the full-degree distribution $p_d(x,y)$ is associated with the chain motifs probability $\rho^c$, and $p_{common}(x,y)$ is associated with the reciprocal motifs probability $\rho^r$. These relationships are expressed as 
\begin{equation}
\begin{aligned}
\langle x^2\rangle&=\int x^2p_{in}(x) dx = \rho^{con}N^2\\
\langle y^2\rangle&=\int y^2p_{out}(y)dy = \rho^{div}N^2\\
\langle xy\rangle &=\int xyp_d(x,y)dxdy= \rho^cN^2\\
\langle \nu\rangle &=\int \nu p_{common}(x,y)d\nu= \rho^rN.
\end{aligned}
\end{equation}
Here $\nu$ represents the number of common nodes in $x$ and $y$.
Therefore, once we define the motif probabilities, we can derive the degree distributions based on the methodology outlined in \cite{chung2002connected,zhao2011synchronization}. 

As for fully-connected Gaussian networks, for simplicity, we assume $\rho^{con}=\rho^{div}=\rho^c$, therefore $\{x_i=y_i\}_{i=1\dots N}$ and  $p_d(x,y)$ has a Gaussian copula with correlation coefficient $1$.

Practically, implementing SONETs requires defining three types of parameters \cite{zhao2011synchronization}. First, the marginal probability of synaptic couplings, denoted by $p$. Second, the probabilities of the reciprocal, convergent, divergent, and chain motifs deviating from independence, denoted by $\alpha_{recip},~\alpha_{conv},~\alpha_{div},~\alpha_{chain}$, respectively. Lastly, the correlation coefficient of the Gaussian copula. All these parameters can be analytically mapped to the parameters we used here to characterize the sparse network with chain motifs:
\begin{equation}
\begin{aligned}
{\rm SONETs}& &{\rm Our\,\,Parameters}\\
p&=&c\\
p^2(1+\alpha_{recip})&=&\rho^r\\
p^2(1+\alpha_{conv})&=&\rho^c\\
p^2(1+\alpha_{div})&=&\rho^c\\
p^2(1+\alpha_{chain})&=&\rho^c
\end{aligned}
\end{equation}
The second and third lines for the convergent and divergent motif statistics are derived by considering the relationship between chains and convergent/divergent motifs (see Appendix.~\ref{ap:GeneralMotifs}), that is, the correlation coefficient of the Gaussian copula is 1.

\section{\label{ap:EffectiveConnectivityMatrix}Effective connectivity matrix for the response function}
In this appendix, we show that the response function obtained from the effective connectivity $\mathbf{J}^{eff} = \mathbf{J^0}+[\mathbf{Z}^2]$ is identical at leading order in $N$ to the mean response $[\boldsymbol{\chi}]$ obtained by averaging  Eq.~\eqref{eq:response_full_rank} over realizations of the random connectivity.

We begin with Eq.~\eqref{eq:response_full_rank}, and use the decomposition $\mathbf{J}=\mathbf{J^0}+\mathbf{Z}$ (Eq.~\eqref{eq:LocallyDefined}). The mean response function $[\boldsymbol{\chi}]$ is expressed as 
\begin{equation}
[\boldsymbol{\chi}] = \left[(\mathbf{1}-\mathbf{Z}-\frac{1}{N}\mathbf{M}_0\mathbf{N}_0^{\intercal})^{-1}\right],
\end{equation}
where the mean connectivity $\mathbf{J^0}=\frac{1}{N}\mathbf{M}_0\mathbf{N}_0^{\intercal}$ has a block structure with constant elements, and is also low-rank.
Using the Woodbury matrix identity and expanding the matrix inverse as a power series \cite{hu2018feedback}, we have

\begin{widetext}
\begin{equation}\label{eq:ap_trialaveraged_response}
    \begin{aligned}
    \left[\boldsymbol{\chi}\right] &= \left[(\mathbf{1}-\mathbf{Z}-\frac{1}{N}\mathbf{M}_0\mathbf{N}_0^{\intercal})^{-1}\right]\\
    &=\left[(\mathbf{1}-\mathbf{Z})^{-1}\right] -\bigg[(\mathbf{1}-\mathbf{Z})^{-1}\left(-\frac{\mathbf{M}_0}{N}\right)\left(\mathbf{1}-{\mathbf{N}_0^{\intercal}(\mathbf{1}-\mathbf{Z})^{-1}\frac{\mathbf{M}_0}{N}}\right)^{-1}\mathbf{N}_0^{\intercal}(\mathbf{1}-\mathbf{Z})^{-1}\bigg]\\
    &\approx\left[(\mathbf{1}-\mathbf{Z})^{-1}\right] +\bigg[(\mathbf{1}-\mathbf{Z})^{-1}\frac{\mathbf{M}_0}{N}\mathbf{N}_0^{\intercal}(\mathbf{1}-\mathbf{Z})^{-1}\bigg]\\
    &\approx\left(\mathbf{1}-\left[\mathbf{Z}^2\right]\right)^{-1} +\bigg[\sum_{k,l=0}^{\infty}\mathbf{Z}^k{\frac{\mathbf{M}_0\mathbf{N}_0^{\intercal}}{N}}\mathbf{Z}^l\bigg]\\
    &=\left(\mathbf{1}-\left[\mathbf{Z}^2\right]\right)^{-1} +\sum_{k,l=0}^{\infty}\left[\mathbf{Z}^2\right]^k\frac{\mathbf{M}_0\mathbf{N}_0^{\intercal}}{N}\left[\mathbf{Z}^2\right]^l.
    \end{aligned}
\end{equation}
\end{widetext}
From the second to the third line, we keep only the first order in $\frac{1}{N}$. From the second-to-last to the last line, we note that the low-rank mean connectivity has a block-like structure with deterministic elements, whereas $\mathbf{Z}$ consists of zero-mean random elements. So $\mathbf{Z}$ is independent of the low-rank mean connectivity, and only terms of the form $[\mathbf{Z}^2]^k$ remain in the large network limit.

We next consider the response function $\boldsymbol{\chi}^{eff}$ generated by $\mathbf{J}^{eff} = \mathbf{J^0}+[\mathbf{Z}^2]$ and perform the same series of manipulations:
\begin{widetext}
\begin{equation}\label{eq:ap_effective_response}
    \begin{aligned}
    \boldsymbol{\chi}^{eff} &= \left(\mathbf{1}-\left[\mathbf{Z}^2\right]-\frac{1}{N}\mathbf{M}_0\mathbf{N}_0^{\intercal}\right)^{-1}\\
    &=\left(\mathbf{1}-\left[\mathbf{Z}^2\right]\right)^{-1} -\left(\mathbf{1}-\left[\mathbf{Z}^2\right]\right)^{-1}\left(-\frac{\mathbf{M}_0}{N}\right)\left(\mathbf{1}-{\mathbf{N}_0^{\intercal}\left(\mathbf{1}-\left[\mathbf{Z}^2\right]\right)^{-1}\frac{\mathbf{M}_0}{N}}\right)^{-1}\mathbf{N}_0^{\intercal}\left(\mathbf{1}-\left[\mathbf{Z}^2\right]\right)^{-1}\\
    &\approx\left(\mathbf{1}-\left[\mathbf{Z}^2\right]\right)^{-1}  +\left(\mathbf{1}-\left[\mathbf{Z}^2\right]\right)^{-1}\frac{\mathbf{M}_0}{N}\mathbf{N}_0^{\intercal}\left(\mathbf{1}-\left[\mathbf{Z}^2\right]\right)^{-1}\\
    &=\left(\mathbf{1}-\left[\mathbf{Z}^2\right]\right)^{-1} +\sum_{k,l=0}^{\infty}\left[\mathbf{Z}^2\right]^k\frac{\mathbf{M}_0\mathbf{N}_0^{\intercal}}{N}\left[\mathbf{Z}^2\right]^l.
    \end{aligned}
\end{equation}
\end{widetext}
Therefore, at leading order in $N$, we find that the averaged response function $[\boldsymbol{\chi}]$ obtained using Eq.~\eqref{eq:ap_trialaveraged_response} is identical to the response function $\boldsymbol{\chi}^{eff}$ derived from the effective connectivity $\mathbf{J}^{eff}=\mathbf{J^0}+[\mathbf{Z}^2]$.

We test the accuracy of $\boldsymbol{\chi}^{eff}$ in both Gaussian (Figs.~\ref{fig:dynamicsExternalInp} and \ref{fig:GaussianParadoxical_nonISN}) and sparse networks (Figs.~\ref{fig:dynamicsExternalInpAdjacency}, \ref{fig:SparsenonISN_InhInhInp} and \ref{fig:betweenGaussian_Sparse} on the right) in the main text. The results show that the response function derived using $\mathbf{J}^{eff}$ is in good agreement with the response function derived using either numerical simulations or those obtained using the low-rank framework.

Examining the eigenvalues of the effective connectivity matrix $\mathbf{J}^{eff}$, we find that they differ from the outliers of the full connectivity matrix $\mathbf{J}$. 
For instance, for fully connected networks with uniform variance (Sec.~\ref{subsecs:EigenvalueGauss}) $\mathbf{J}^{eff}$ has a  unit rank structure, and therefore only one non-zero eigenvalue, while $\mathbf{J}$ has two outliers as $\tau^c$ is increased (Fig.~\ref{fig:F1_ConnStatsOutliers}). The single unique eigenvalue is given by 
\begin{equation}
\lambda_{eff} = \lambda_0 + N \sigma^2 \tau^c,
\end{equation}
and therefore increases linearly with $\tau^c$, while the positive outlier  of $\mathbf{J}$ increases as $\sqrt{\tau^c}$ (Eqs.~(\ref{eq:twoOutliersMaintext}-\ref{eq:delta_chain})).

Despite this difference, we find that $\mathbf{J}^{eff}$ predicts correctly network instability. Indeed for the full connectivity matrix $\mathbf{J}$, the positive outlier crosses unity when 
\begin{equation}
\frac{\lambda_0+\sqrt{\lambda_0^2+4(N-1) \sigma^2 \tau^c}}{2}=1,
\end{equation}
here we use $N$ instead of $N-1$ (Eqs.~\eqref{eq:twoOutliersMaintext},~\eqref{eq:delta_chain}) in the large network limit.
Rearranging terms, this is directly equivalent to $\lambda_{eff} =1$.
Therefore, the effective connectivity matrix $\mathbf{J}^{eff}$ accurately predicts the instability of the network dynamics.

\section{\label{ap:figures}Supplementary figures.}

\begin{figure*}[htb]
\includegraphics[width=0.98\textwidth]{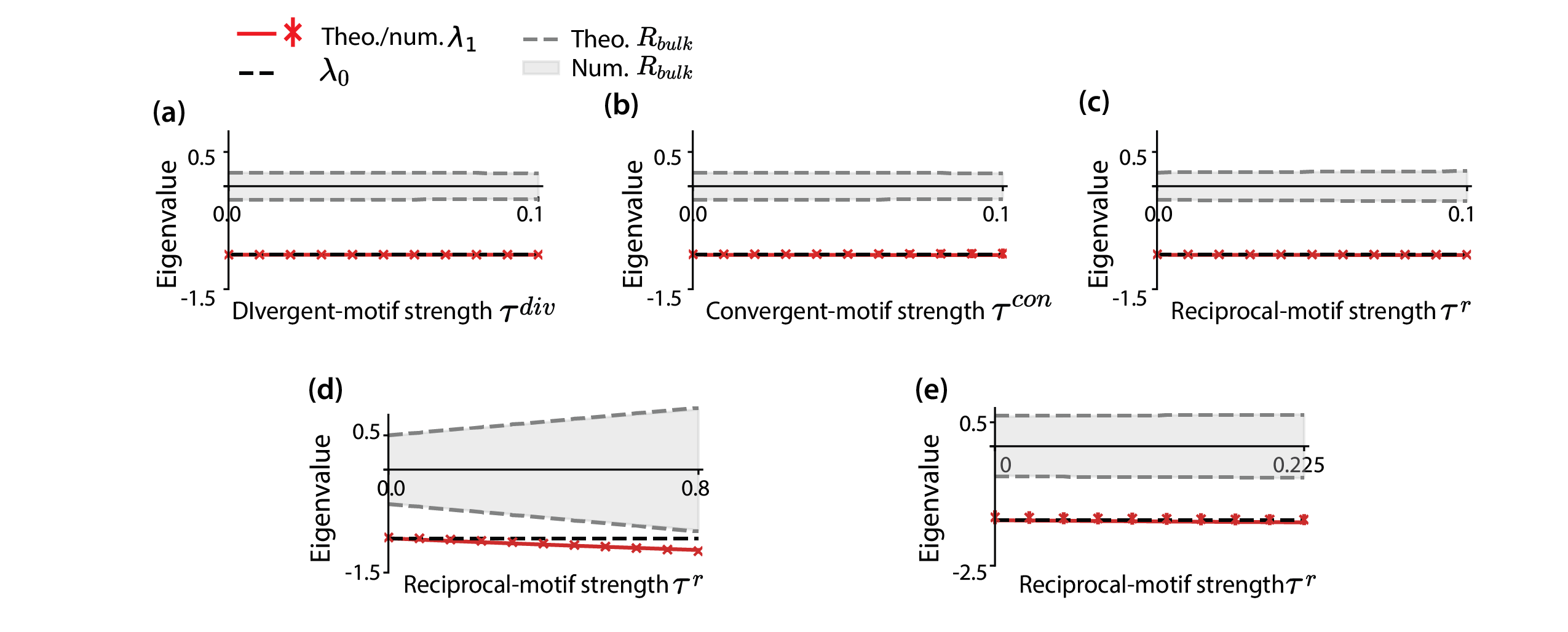}
\caption{\label{fig:comparable_ReciprocalMotifs} Eigenvalues of networks featuring divergent, convergent and reciprocal motifs.
In subplots (a-d), results are presented for networks with Gaussian-distributed synaptic couplings. Subplots (a), (b), (c) and (d) show how the unique eigenvalue outlier $\lambda_1$ and the radius of the eigenvalue bulk $R_{bulk}$ change as the statistics of the divergent $\tau^{div}$, convergent $\tau^{con}$ and reciprocal motifs $\tau^r$ increase. The red error bars with asterisks represent the numerically obtained $\lambda_1$ from 30 network realizations, the red lines represent the theoretical predictions. 
Gray areas illustrate the area of eigenvalue bulks obtained numerically, with gray dashed lines representing the theoretical predictions \cite{dahmen2020strong}. The black dashed lines indicate the unperturbed eigenvalue $\lambda_0$ of the mean connectivity. In subplot (a-c), the network parameters are the same as those used in Fig~\ref{fig:F1_ConnStatsOutliers}: $\sigma=0.2$ and $\tau^{div/con/r}$ ranges from $0$ to $0.1$. In subplot (d), the networks have parameters $\sigma=0.5$, and $\tau^r$ ranges from $0$ to $0.8$.
Subplot (e) exhibits results for sparse networks. Figure descriptions are consistent with those in subplots (a-d).}
\end{figure*}

\begin{figure*}[htb]
\includegraphics[width=0.98\textwidth]{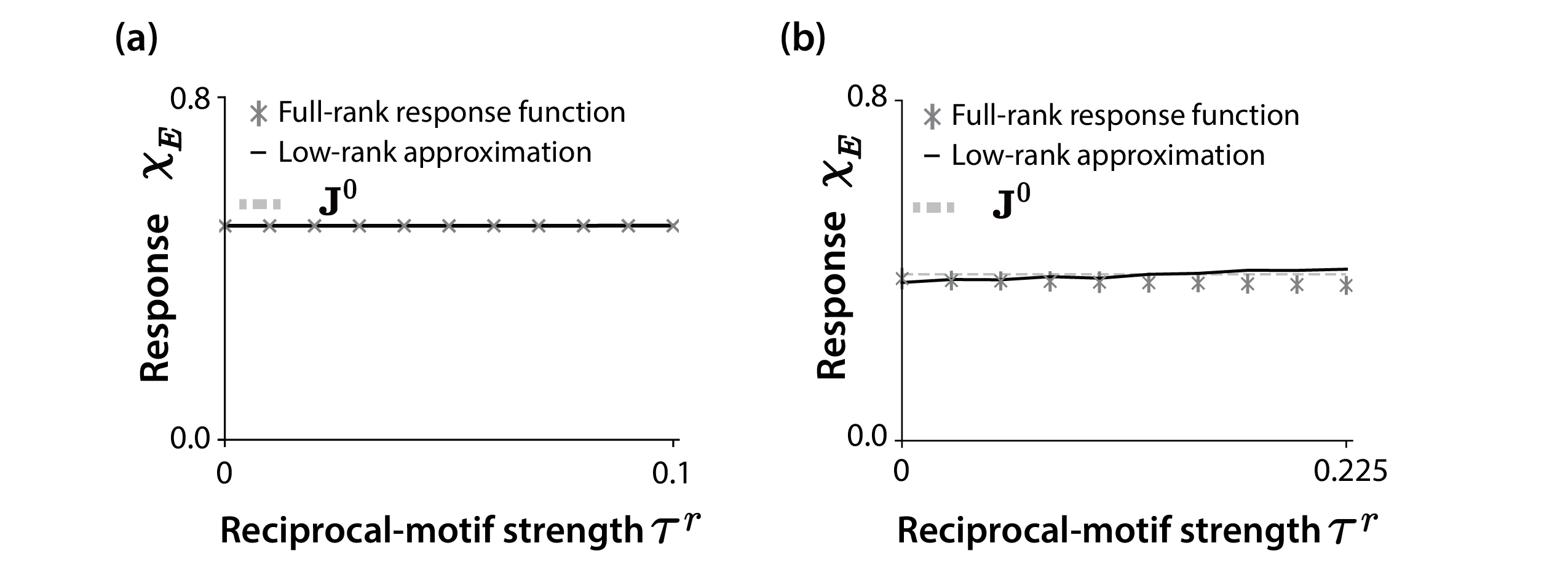}
\caption{\label{fig:ReciprocalDyns}
Mean response $\chi_E$ of the excitatory neuron population in response to the time-invariant uniform input $I^{ext}$ in networks with only reciprocal motifs. Subplots (a) corresponds to Gaussian networks, while subplots (b) corresponds to sparse networks. Network parameters remain consistent with those provided in Figs~\ref{fig:dynamicsExternalInp} and \ref{fig:dynamicsExternalInpAdjacency}.
}
\end{figure*}

\begin{figure*}[htb]
\includegraphics[width=0.98\textwidth]{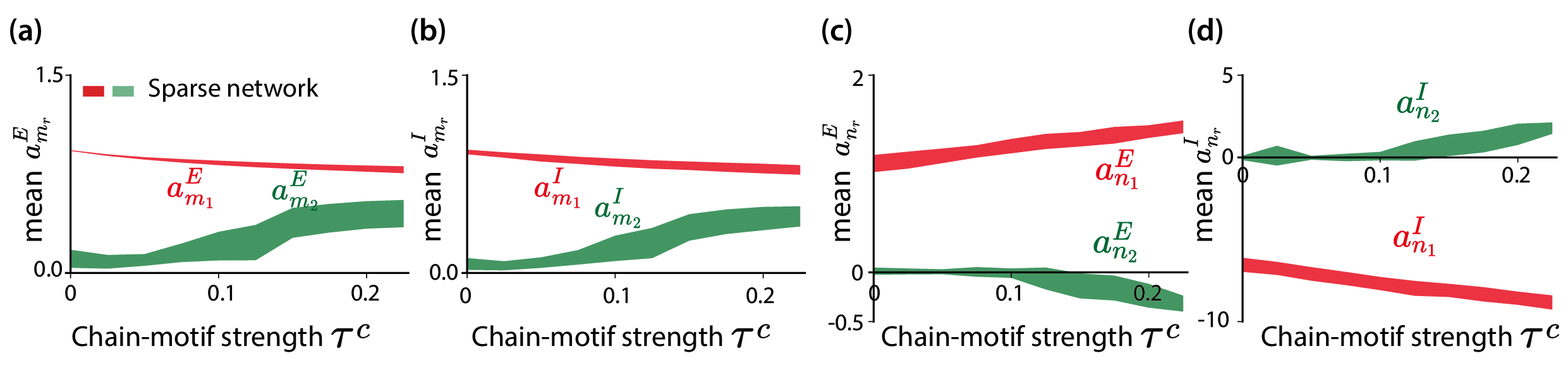}
\caption{\label{fig:Supp_AdjacencymatrixVectors}
Population-averaged mean of entries on connectivity vectors for sparse networks featuring chain motifs.
Figure descriptions are identical to those in Fig.~\ref{fig:AdjacencymatrixVectors}, except that the network parameters are consistent with those used in Fig.~\ref{fig:SparsenonISN_InhInhInp} (see TABLE~\ref{tab:parameters}).}
\end{figure*}

\begin{figure*}[htb]
\includegraphics[width=0.88\textwidth]{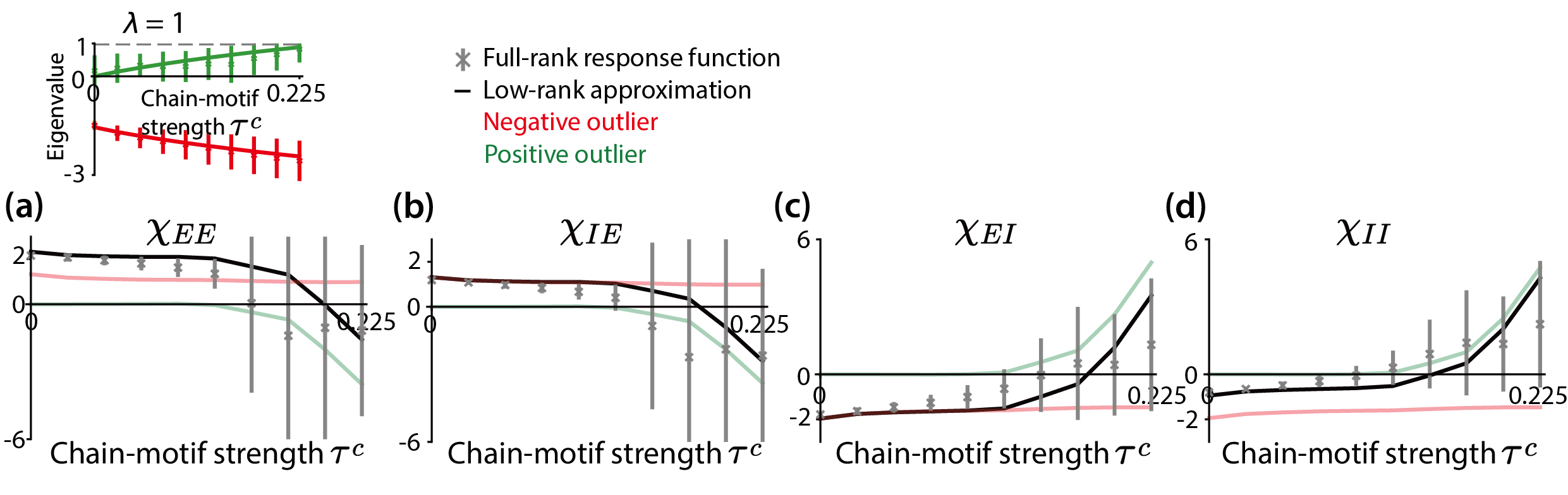}
\caption{\label{fig:SparseISN_2_NonISN}
Impact of chain motifs on the mean response $\chi_{pq}$ of population $p$ to uniform inputs to population $q$, in  sparse excitatory-inhibitory networks. Figure descriptions are identical to those in Fig.~\ref{fig:SparsenonISN_InhInhInp}, except that the network parameters are consistent with those used in Figs~\ref{fig:AdjacencymatrixVectors},~\ref{fig:dynamicsExternalInpAdjacency} (see TABLE~\ref{tab:parameters}).}
\end{figure*}

\begin{figure*}[htb]
\includegraphics[width=0.92\textwidth]{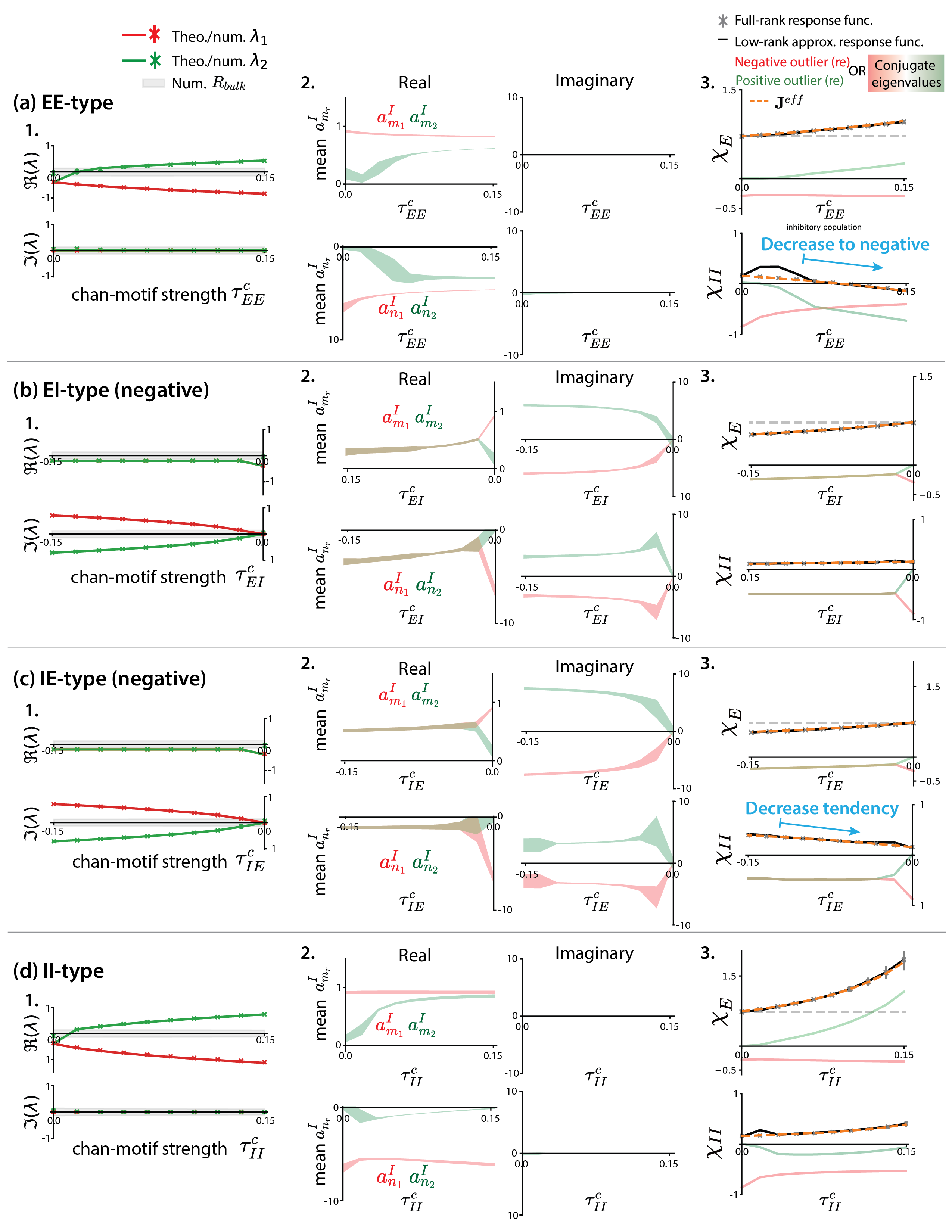}
\caption{\label{fig:Supp_celltype_contributions} Distinct effects of EE-type (panel (a), $\tau^c_{EE}>0$), EI-type (panel (b), $\tau^c_{EI}<0$), IE-type (panel (c), $\tau^c_{IE}<0$), and II-type (panel (d), $\tau^c_{II}$) chain motifs on eigenvalues (1.), inhibitory-associated low-rank structures (2.), and response functions $\chi_{E}$ and $\chi_{II}$ (3.). Results are obtained using the Gaussian approximation for sparse networks; except for the varying chain-motif strength, network parameters are identical to those in Figs.\ref{fig:SparsenonISN_InhInhInp} and\ref{fig:betweenGaussian_Sparse}.}
\end{figure*}

\begin{figure*}[htb]
\includegraphics[width=0.92\textwidth]{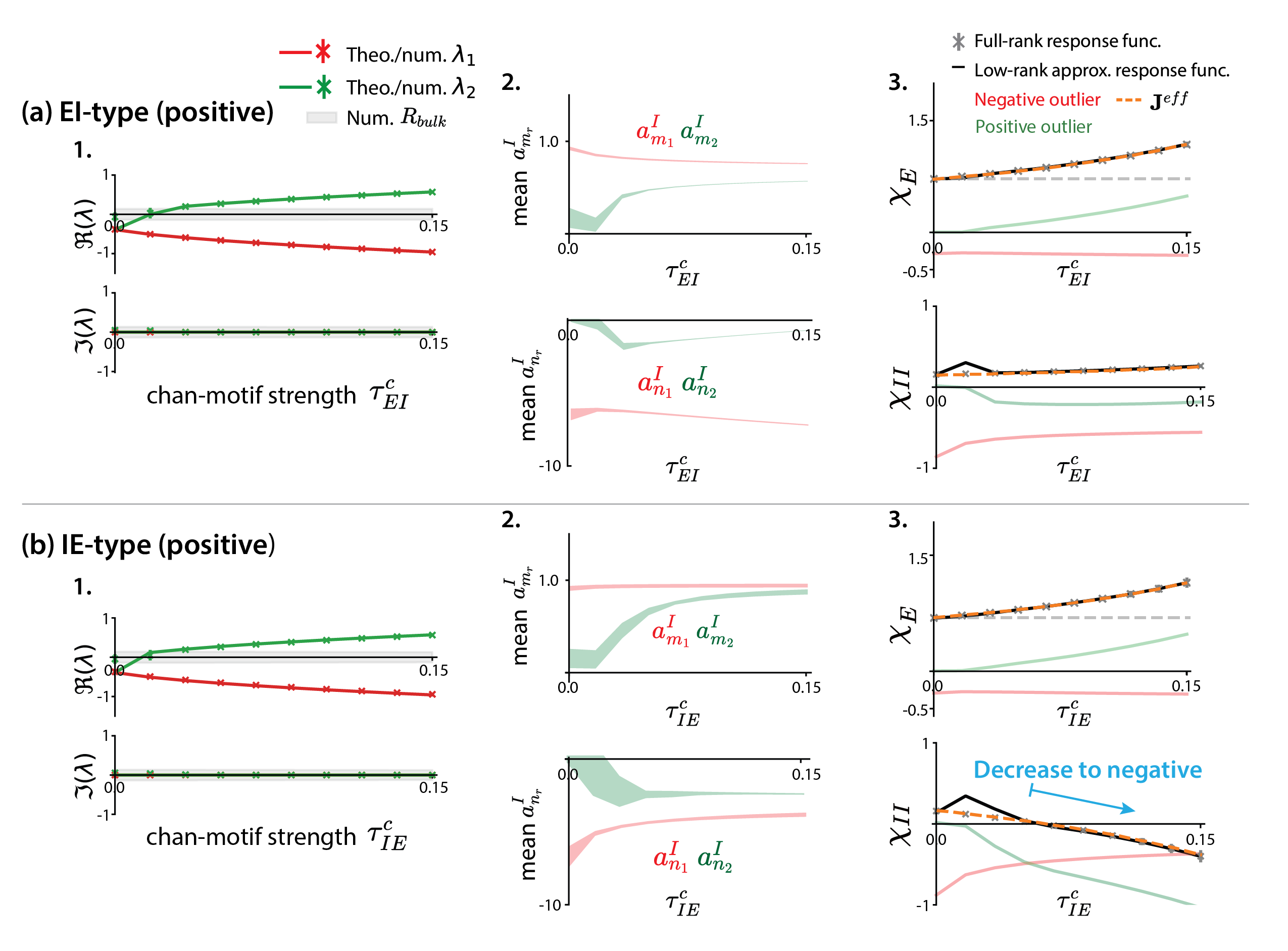}
\caption{\label{fig:Supp_celltype_neg_contribution}
Distinct effects of EI-type (panel (a), $\tau^c_{EI}>0$) and IE-type (panel (b), $\tau^c_{IE}>0$) chain motifs on eigenvalues (1.), inhibitory-associated low-rank structures (2.), and response functions $\chi_{E}$ and $\chi_{II}$ (3.). Other figure descriptions remain consistent with those in Fig.~\ref{fig:Supp_celltype_contributions}.}
\end{figure*}

\end{document}